\documentclass[aps, prx, amsmath, amssymb, a4paper, floatfix, superscriptaddress, longbibliography]{revtex4-2}

\usepackage{ulem}
\usepackage{algorithm}
\usepackage{algorithmicx}
\usepackage{algpseudocode}
\usepackage{amsfonts}
\usepackage{amsthm}
\usepackage[title]{appendix}
\usepackage[english]{babel}
\usepackage{bm}
\usepackage{booktabs}
\usepackage{comment}
\usepackage[utf8x]{inputenc} 
\usepackage{csquotes}
\usepackage{graphicx} 
\usepackage{listings}
\usepackage{mathrsfs}
\usepackage[version=4]{mhchem}
\usepackage{multirow}
\usepackage{physics}
\usepackage{siunitx}
\usepackage{textcomp}
\usepackage{textgreek}
\usepackage{xcolor}
\begin{document}

\title[]{Measurement of traveling pressure waves inside a droplet}

\author{Sayaka Ichihara}
\email{sayaka.ichihara@ovgu.de}
\affiliation{Department of Mechanical Systems Engineering, Tokyo University of Agriculture and Technology, Nakacho 2-24-16, Koganei, 184-8588, Tokyo, Japan}
\affiliation{Faculty of Natural Sciences, Institute for Physics, Otto-von-Guericke-University Magdeburg, Universitätsplatz 2, 39106, Magdeburg, Germany}

\author{Samuele Fiorini} 

\affiliation{Institute of Fluid Dynamics, D-MAVT, ETH Zürich, Sonneggstrasse 3, Zürich, 8092, Switzerland}

\author{Yoshiyuki Tagawa}
\email{tagawayo@cc.tuat.ac.jp}
\affiliation{Department of Mechanical Systems Engineering, Tokyo University of Agriculture and Technology, Nakacho 2-24-16, Koganei, 184-8588, Tokyo, Japan}

\author{Outi Supponen}
\affiliation{Institute of Fluid Dynamics, D-MAVT, ETH Zürich, Sonneggstrasse 3, Zürich, 8092, Switzerland}

\begin{abstract}
Shock wave--droplet interactions have been receiving increasing attention due to their relevance in aviation fuel combustion and minimally invasive medical treatments, yet quantifying them experimentally remains a challenge.
In this study, we propose a background-oriented schlieren (BOS) technique for quantitative spatiotemporal measurements of shock wave-droplet interaction, employing a novel ray-tracing correction, a synchronization system, and a projected background. 
Underwater shock waves propagating both inside and outside a millimetric perfluorohexane droplet immersed in water are experimentally measured. 
The quantified density-gradient and pressure fields are compared with numerical simulations, and the BOS measurements—including sound speeds, the shock-focusing location, and the maximum pressure—are found to be in close agreement with the numerical results. 
Notably, the technique successfully captures the phase shift before and after shock focusing that had previously only been hypothesized. 
\end{abstract}

\maketitle
\section{Introduction}\label{sec:intro}
%-------------
Shock wave-droplet interactions have been extensively studied for their many practical applications, including agricultural sprays \cite{shinjo2011surface,sharma2021shock}, aircraft propulsion systems \cite{hsiang1995drop,tripathi2022interactions}, rainfall phenomena \cite{field1989effects,forehand2023numerical}, and drug delivery \cite{menezes2009shock}. 
 When a shock wave interacts with a droplet, depending on its strength, deformation and breakup may occur due to viscous drag, Rayleigh–Taylor instability, or shear \cite{theofanous2004aerobreakup}.
Recent studies indicate that, prior to droplet deformation, shock waves within the droplet can generate tensile stresses that induce phase transitions from liquid to gas \cite{sembian2016plane,fiorini2025positive}. 
However, quantifying the internal dynamics poses technical challenges for both numerical and experimental approaches.
 Igra \textit{et al.} demonstrated similarities between the breakup dynamics of a cylindrical water column and a spherical droplet driven by a shock wave \cite{igra2010shock}.
The water column configuration has therefore been widely adopted to study shock wave–droplet interactions, as it allows easy control of the droplet radius and position, provides optical access, and enables a valid two-dimensional approximation.
As a result, both experimental and numerical studies have been able to expand their focus beyond droplet deformation to internal phenomena, including shock-wave propagation within the droplet.

Numerical simulations offer the advantage of resolving shock-wave propagation and droplet deformation in detail \cite{igra2010shock}.
However, their applicability often relies on fitted parameters derived from quantitative measurements \cite{tripathi2022interactions, forehand2023numerical, damianos2025effect} and becomes increasingly challenging and computationally expensive when phase transitions and complex non-homogeneous media are involved \cite{xiong2024exploration}.
Quantitative measurement results are required to support the limitations of numerical simulations \cite{forehand2023numerical,haas1987interaction}.
Non-contact visualization techniques, including Schlieren imaging \cite{field1989effects}, shadowgraphy \cite{sembian2016plane}, and holographic interferometry \cite{igra2001investigation,igra2003experimental}, have been widely employed to visualize shock-wave propagation and droplet deformation.
While these methods provide image intensity related to first- or second-order density variations, quantitative evaluation generally requires complex calibration procedures, and fully quantitative measurements remain limited \cite{settles2001schlieren}.
To address these challenges, a quantitative approach based on schlieren imaging combined with piezoelectric sensors in a water column was proposed \cite{sembian2016plane}.
However, contact-based sensors disturb the flow field, making their application to droplets difficult.
%-----------------------------------
\begin{figure}
    \centering
    \includegraphics[width=0.5\linewidth]{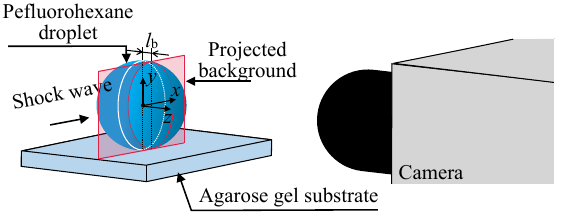}
    \caption{Schematic of the experimental setup for the proposed BOS technique for quantitative measurements inside a droplet. 
        A perfluorohexane (PFH) droplet is placed on an agarose gel substrate in water to stabilize its position. 
        The origin of the ($x,y,z$) coordinate system is set at the center of the droplet.
        A shock wave propagates along the $x$-axis from left to right. 
        A background pattern is projected and positioned at $z = l_b$ along the optical axis, and the camera is focused on the background. 
    }
    \label{fig:droplet-BOS}
\end{figure}
%-----------------------------------

Particle image velocimetry (PIV) has been successfully applied to measure velocity fields inside and around droplets in microchannels \cite{kinoshita2007three,liu2017micro,miessner2020mupiv}, evaporating droplets \cite{kang2004quantitative}, and acoustically levitated droplets \cite{yamamoto2008internal}.
A major challenge in non-contact measurements of shock wave–droplet interactions is optical distortion arising from the refractive-index difference between the droplet and the surrounding fluid.
In PIV, this distortion has been corrected using ray-tracing corrections \cite{kang2004quantitative,minor2007optical,kumar2017internal} and index-matching techniques \cite{timgren2008application}.
Another challenge is capturing pressure waves exceeding several MPa in amplitude and propagating at the speed of sound.
PIV is unsuitable in this regime because tracer particles cannot follow the shock waves, and the required temporal resolution is difficult to achieve \cite{miessner2020mupiv}.

Background-oriented schlieren (BOS) is a simple optical technique based on the relationship between refractive index and density \cite{venkatakrishnan2004density,raffel2015background}.
It measures the distortion of a background pattern induced by refractive-index variations, which is proportional to the line-of-sight–integrated density gradient \cite{meier2002computerized,venkatakrishnan2004density}.
BOS does not require calibration experiments, as it relies solely on the distortion of the background pattern.
Using a projected background instead of a physical one, quantitative measurements can be performed without disturbing the flow \cite{leopold2012increase,ota2015improvement}.
Moreover, the high spatiotemporal resolution required for quantitative shock measurements can be achieved by applying a synchronization technique \cite{ichihara2025high,hayasaka2016optical}.

In this study, we propose background-oriented schlieren with a projected background as an easy-to-implement, quantitative and non-invasive measurement technique for shock waves propagating inside a spherical droplet.  %perfluorohexane (\ce{C6F14}, PFH) 
An illustrative schematic of the experimental setup is shown in Fig.~\ref{fig:droplet-BOS}. 
Optical distortions caused by refractive-index difference are taken into account by applying a proposed ray-tracing correction. 
We utilize an enhanced BOS technique developed by the authors—fast checkerboard demodulation (FCD) and vector tomography-BOS (VT-BOS).
These methods provide higher spatial resolution and improved measurement accuracy compared to conventional BOS, making them particularly suitable for shock wave measurements \cite{shimazaki2022background,ichihara2022background}. 
The experimentally quantified, spatio-temporally resolved integrated density gradient is compared with numerical simulation results to evaluate shock-wave propagation characteristics, including its distribution, propagation speed, and focusing position. % \textcolor{red}{and the discrepancies between them.} 
On the other hand, the measured pressure field, including its uncertainty, is compared with numerical simulations to evaluate the measurement accuracy.

%%===============================================
\section{Measurement principle}
\subsection{Background-oriented schlieren}\label{sec:bos}
%----------------------------
\begin{figure}[ht]
    \centering
    \includegraphics[width=0.5\linewidth]{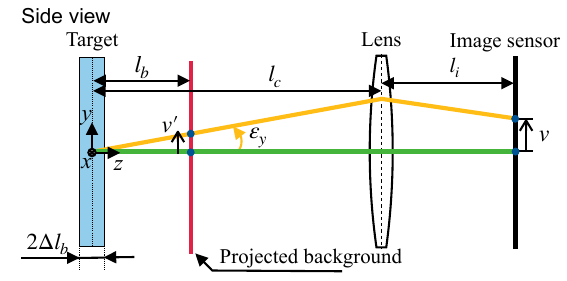}
    \caption{A schematic of the BOS theory.}
    \label{fig:bos}
\end{figure}
%---------------------------

BOS is a non-contact quantitative measurement technique that only requires a background and a camera \cite{raffel2015background}.
The principle of BOS relies on the relationship between the refractive index $n$ and the density $\rho$, which is expressed by the Gladstone–Dale equation: 
%------------------------------------
\begin{gather}
    n = \rho K + 1, 
    \label{eq:bos_gland-stone}
\end{gather}
%------------------------------------
 where $K$ is the Gladstone–Dale constant \cite{venkatakrishnan2004density,venkatakrishnan2005density}.

A typical optical alignment of BOS on the $y$-$z$ plane is illustrated in Fig.~\ref{fig:bos}.  
The image sensor, camera lens, projected background, and target are positioned along the optical axis of the camera.  
The background is projected into the fluid using two lenses \cite{leopold2012increase}.
The camera’s focal plane is aligned with the image plane of the projected background.
Herein the distance between the two lenses and the background determines the position of the projected image, whereas the focal length of the two lens governs the projection magnification.
The green line represents a light ray that passes through the center of the background and is captured by the sensor when the target is not present in the optical path, while the yellow line represents the same light ray refracted by the target.
The deviation angle $\boldsymbol{\varepsilon}$ between the yellow and green lines contains information about the integrated gradient of the refractive index along the optical axis \cite{venkatakrishnan2004density}:
%------------------------------------
\begin{align}
    \boldsymbol{\varepsilon} = \left( \varepsilon_x,\varepsilon_y \right) 
    &=  \frac{1}{n_0} \int^{l_b + \Delta l_b}_{l_b - \Delta l_b} \frac{\partial n}{\partial \boldsymbol{r}} dz \notag \\
    &= \frac{1}{n_0} \int^{l_b + \Delta l_b}_{l_b - \Delta l_b} \left( \frac{\partial n}{\partial x}, \frac{\partial n}{\partial y} \right) dz,
    \label{eq:bos_epsilon}
\end{align}
%------------------------------------
where $n_0$ is the refractive index of the surrounding fluid, $l_b$ is the distance from the center of the target to the background, $2\Delta l_b$ is the thickness of the target, the vector $\boldsymbol{r}$ represents the position in Cartesian coordinates $(x,y)$, and the origin $O$ is at the center of the measurement target.

The deviation angle $\varepsilon_y$ along the $y$-axis on the $y$–$z$ plane appears as the displacement component $v$ on the image sensor.
Similarly, the deviation angle $\varepsilon_x$ along the $x$-axis on the $x$–$z$ plane corresponds to the displacement component $u$.
Accordingly, the overall vector field can be expressed as $\boldsymbol{w} = (u, v)$.
These displacements represent distortions of the background pattern, which is located in the focus of the camera.
In this paper, the order of magnitude of $l_b$ is the same as that of $2\Delta l_b$.
Under this condition, the relationship between displacement and deviation angle satisfies the near-field BOS principle,
%------------------------------------
\begin{align}
    \boldsymbol{w} = M \frac{(l_b + \Delta l_b)^2}{l_b} \boldsymbol{\varepsilon},
    \label{eq:bos_disp_epsilon} 
\end{align}
%------------------------------------
where $M=l_i/(l_b + l_c)$ is the camera's magnification, $l_i$ is the distance between lens and image sensor, $l_c$ is the distance between projected background and lens, and $(l_b+\Delta l_b)^2 \boldsymbol{\varepsilon} /l_b $ represents the amount of distortion in the background \cite{van2014density,raffel2015background}.
The relationship between displacement and refractive index is obtained by substituting Eqs.~\eqref{eq:bos_gland-stone} and \eqref{eq:bos_epsilon} into Eq.~\eqref{eq:bos_disp_epsilon},
%------------------------------------
\begin{align}
    \boldsymbol{w} &= \frac{M}{n_0}\frac{(l_b + \Delta l_b)^2}{l_b}\int \frac{\partial n}{\partial \boldsymbol{r}} dz \\
    &= \frac{K M}{n_0}\frac{(l_b + \Delta l_b)^2}{l_b}\int \frac{\partial \rho}{\partial \boldsymbol{r}} dz.
    \label{eq:bos_disp}
\end{align}
%------------------------------------
The displacement $\boldsymbol{w}$ is obtained by comparing the distorted image (with target) and a reference image (without target) using a displacement detection technique, such as cross-correlation method \cite{westerweel1997fundamentals} or optical flow \cite{atcheson2009evaluation,hayasaka2016optical}.
This study employs fast checkerboard demodulation (FCD) \cite{wildeman2018real} and a checker background pattern to measure a density gradient field.
FCD is well suited for measuring high-density-gradient fields (e.g., shock waves and high-amplitude ultrasound) because it preserves high spatial resolution by using a Fast Fourier Transform, unlike cross-correlation-based methods. 
It has also been reported that FCD can measure density gradients up to approximately 2.5 times larger than those accessible with PIV or optical flow \cite{shimazaki2022background, ichihara2025high}.

After calculating the displacement, three-dimensional reconstruction is applied to the integrated density gradient field to estimate the reconstructed field (non-uniform density gradient field) along the $z$-axis \cite{venkatakrishnan2004density}.
This study uses vector tomography (VT) as the reconstruction technique, which is based on two key assumptions. 
The first assumption is that the reconstructed field, expressed in a Cartesian coordinate centered at the reconstructed field, is symmetric about an axis perpendicular to the projection direction (the $z$-axis), such as the $x$- or $y$-axis. 
The second assumption is that the reconstructed field, expressed in a polar coordinate centered at the reconstructed field, is governed solely by the radial component. 
Under these assumptions, VT can directly reconstruct the three-dimensional distribution from the integrated values of the vector field using an inverse matrix. 
Compared with filtered back projection (FBP), VT reduces computational cost and mitigates the loss of reconstruction accuracy associated with limited spatial resolution \cite{ichihara2022background}.

The density field obtained by integrating $\partial \rho / \partial x$ along the $x$-axis is related to the pressure through the equation of state (EoS), specifically the Noble--Abel stiffened-gas equation of state (NASG--EoS), given by
%------------------------------------
\begin{align}
    p = \frac{\rho(\gamma - 1) C_v T_0}{(1 - b\rho)} - P_\infty,
    \label{eq:bos_nasg-eos}
\end{align}
%------------------------------------
where $\gamma$ is the specific heat ratio, $C_v$ is the specific heat at constant volume, $T_0$ is the temperature estimated from the NASG--EoS (Eq. \ref{eq:bos_nasg-eos}) at atmospheric pressure $p_0$, $b$ is the parameter representing repulsive intermolecular effects of the material, and $P_\infty$ is the stiffness parameter of the surrounding fluid \cite{le2016noble}.  
The values of these constants were adopted from Prasanna \textit{et al.,} who applied the NASG--EoS to perfluorohexane (PFH) \cite{Prasanna2025PFHEquationOfState}.

%%===============================================
%--------------------------------------------
\begin{figure}[ht]
    \centering
    \includegraphics[width=0.5\linewidth]{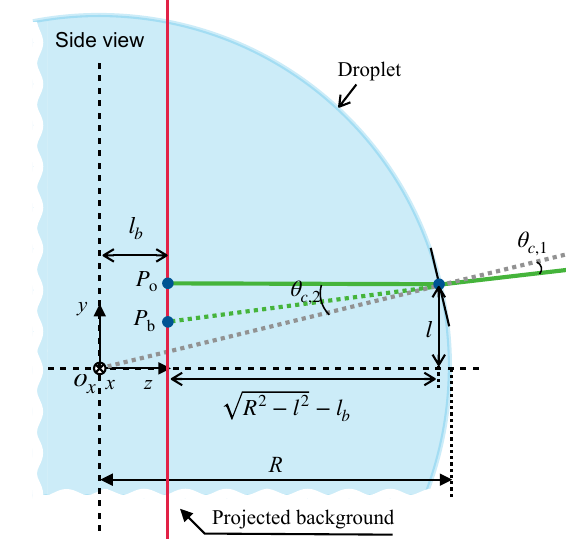}
    \caption{Schematic illustrating the optical path distortion considered in the coordinate correction.
    A light ray passing through point $P_0$ is refracted by the droplet on the $y$–$z$ plane, as shown by the green line.
    The dashed green line indicates the corresponding ray imaged onto the camera sensor, which is displaced due to refraction at the droplet interface.
    The red line represents the projected background.}
    \label{fig:raytrace-theory-coordinate}
\end{figure}

\begin{figure}[ht]
    \centering
    \includegraphics[width=1.0\linewidth]{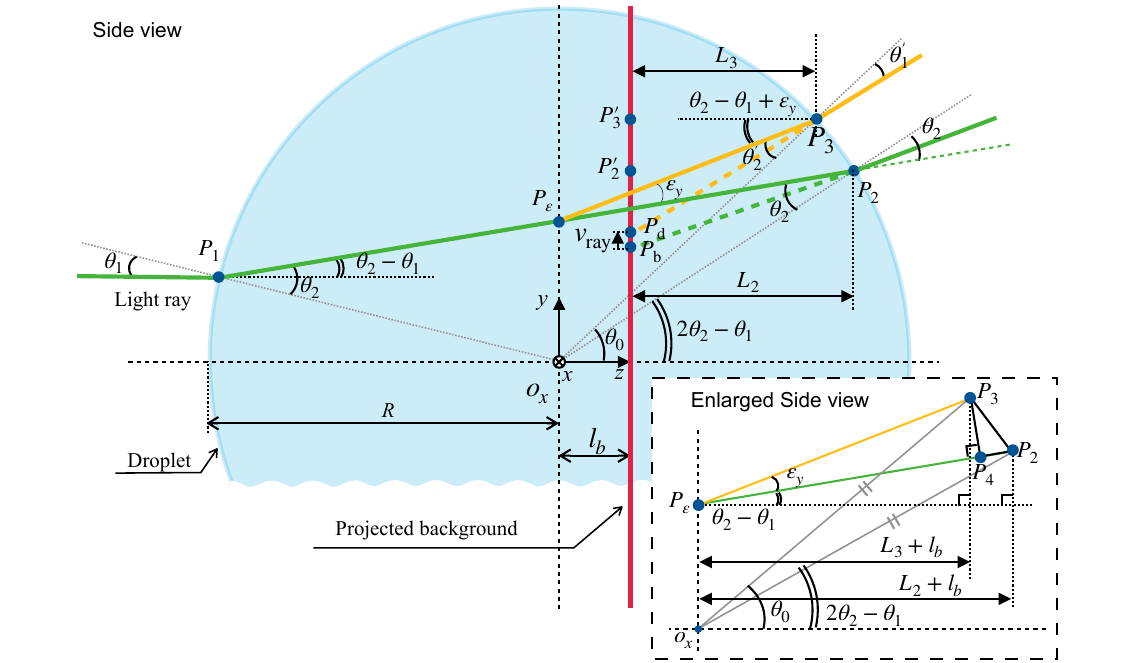}
    \caption{
    The geometry of light rays refracted by the droplet on the $y$–$z$ cross-section.
    The refraction angles in the surrounding fluid and within the droplet are denoted as $ \theta_1 $ and $ \theta_2 $, respectively.  
    $ R$ represents the droplet radius.
    $ l_b $ denotes the distance from the droplet center to the projected background. 
    The yellow and green lines represent the light rays passing through the droplet in the presence and absence of the measurement target, respectively. }
    \label{fig:raytrace-theory-displacement}
\end{figure}
%--------------------------------------------
%********************************************************
\subsection{Ray tracing for a spherical object}\label{sec:raytracing}

This section proposes a ray tracing technique to correct the refraction of light rays passing through a spherical droplet.
The ray tracing is known as tracing back a light ray imaged onto a sensor \cite{glassner1989introduction}.
This technique has been widely applied in optical measurement methods \cite{kang2004quantitative}.
% To the best of our knowledge, this study is the first to introduce ray tracing for correcting refraction in a spherical object within the BOS framework.
Here, ray tracing is introduced within the BOS framework.
%-----------------------------------------
\subsubsection{Coordinate correction}
This subsection presents a correction method to compensate for the spatial misalignment of the coordinate system induced by refraction at the curved droplet interface. 
The displacements $(u,v)$ are treated as independent constants.
Consequently, in the coordinate and displacement corrections along the $x$- and $y$-directions, the rays in each direction can be considered independently.

We consider the coordinate correction in the $y$-direction.
Provided that the droplet size and the lens are much larger than the wavelength of light, propagation of light rays can be treated as geometrical optics \cite{kang2004quantitative,hecht2002optics}.
The light-blue disk represents the spherical droplet on the $y$-$z$ plane at a given $x$-coordinate, with radius $R$.
The origin $O_x$ is located at the same $x$-coordinate along the $x$-axis.  
The red solid line denotes the projected background shown in Fig.~\ref{fig:raytrace-theory-coordinate}.  
A light ray entering through point $P_o$, parallel to the $z$-axis, is refracted at the droplet interface and subsequently enters the surrounding fluid.  
Consequently, the ray deviates from its original position $P_o$ on the projected background and is recorded instead at position $P_b$.  
To correct for this deviation, we define a coordinate correction vector $\overrightarrow{P_oP_b}$, given by
\begin{equation}
    \overrightarrow{P_oP_b} = \left(\sqrt{R^2 - l^2} - l_b \right) \tan{(\theta_{c,2} - \theta_{c,1})},
    \label{eq:ray-PoPb}
\end{equation}
where $l$ is the distance from the origin $O_x$ to point $P_o$, and $\theta_{c,1}$ and $\theta_{c,2}$ are the refraction angles at the droplet interface.  
By estimating this $y$-coordinate correction at points along $-R_{\rm drop} < x < R_{\rm drop}$, the coordinate correction in the $y$-direction can be applied to all points on the $x$-$y$ plane.  
Here, $R_{\rm drop}$ denotes the droplet radius on the $y$-$z$ plane at $x=0$.

The coordinate correction in the $x$-direction can be derived in a manner analogous to that in the $y$-direction by considering the light ray geometry on the $x$-$z$ plane on a given $y$-coordinate. 
Since the $x$- and $y$-direction displacements are independent, the $x$-coordinate correction is equivalent to the result obtained by rotating the $y$-coordinate correction on the $x$-$y$ plane by 90 $^{\circ}$ clockwise about the origin $O$.

%********************************************************
\subsubsection{Displacement correction}
The geometry of light rays refracted by the droplet on the $y$-$z$ cross-section at a given $x$-coordinate is illustrated in Fig.~\ref{fig:raytrace-theory-displacement}.
The subscripts 1 and 2 denote the surrounding fluid and the droplet, respectively.
The measurement target is assumed to locate on the $x$-$y$ plane passing through the origin $O$. 
The near-field BOS theory accounts for the effect of the target's thickness on the geometry of the light rays passing through it and provides a linear relation between the displacement and the deviation angle, as given in Eq.~\eqref{eq:bos_disp_epsilon}.
Based on Eq.~\eqref{eq:bos_disp_epsilon}, the effect of the target thickness on the geometry of the light rays passing through a target located on the $y$-axis is neglected.
The incident light rays at the droplet are assumed to be parallel along the optical axis ($z$-axis).

We only describe the $y$-direction displacement correction.
The corrections in the $x$- and $y$-directions are independent of each other, and can be treated in an analogous manner.
The yellow and green lines represent the light rays passing through the droplet in the presence and absence of the measurement target, respectively.
Specifically, the light ray passing through point $P_{\varepsilon}$ is refracted by a deviation angle $\varepsilon_y$ due to density gradient induced by the measurement target (e.g., a shock wave) located along the $y$-axis.
This light ray subsequently enters the surrounding fluid at point $P_3$.
In contrast, when the measurement target is absent, the light ray passing through point $P_{\varepsilon}$ enters the surrounding fluid at point $P_2$.
Here, the points $P'_2$ and $P'_3$ represent the positions obtained by projecting the points $P_2$ and $P_3$ onto the plane of projected-background, respectively.

In BOS technique, the camera is focused on the projected background pattern. 
The points $P_b$ and $P_d$ denote the apparent image positions of the green and yellow light rays on the image sensor.
The vector $\overrightarrow{P_bP_d}$ represents the displacement $v_{\mathrm{ray}}$ obtained from the reference and distorted images.
The light rays passing through points $P_2$ and $P_3$, corresponding to the green and yellow rays, are influenced by both the density field of the target and the refractive index distribution of the droplet.
Unlike the $y$-direction displacement $v$ shown in Eq.~\eqref{eq:bos_disp_epsilon}, the displacement $v_{\mathrm{ray}}$ does not uniquely determine the deviation angle $\varepsilon_y$.
Therefore, we propose a displacement correction that estimates the deviation angle $\varepsilon_y$ from displacement $v_{\mathrm{ray}}$ using $R$, $n_1$, $n_2$, $\theta_1$ and $\theta_2$.

According to Snell’s law, the angle of incidence $\theta'_1$ and angle of refraction $\theta'_2$ at point $P_3$ is given by
%---------------
\begin{equation}
    n_1 \sin{\theta'_1} = n_2 \sin{\theta'_2}.
    \label{eq:snell}
\end{equation}
%-----------------------------
Let $\theta_0$ denote the angle of the line $O_xP_3$ intersecting the $z$-axis.
Using the concept of alternate interior angles with respect to the line parallel to the $z$-axis passing through point $P_3$, $\theta_0$ can be expressed as follows:
%-------------------------------
\begin{equation}
    \theta_{\rm 0} = \theta_2 - \theta_1 +\varepsilon_y + \theta'_{2}.
    \label{eq:theta0}
\end{equation}
%-----------------------------
The distances between the projected background and the points $P_2$ and $P_3$ along the $z$-axis are defined as $L_2 = R  \cos{\left( 2\theta_2-\theta_1\right)} - l_b$, 
$L_3 = R \cos{\theta_0} - l_b$.
The displacement vector $v_{\mathrm{ray}}$ is describes as follows:
%-------------------------------
\begin{equation}
\begin{split}
    v_{\mathrm{ray}} &=\overrightarrow{P_bP_d} 
    = \overrightarrow{P'_3P_d} - \overrightarrow{P'_3P'_2} - \overrightarrow{P'_2P_b}
    \\
    &=L_3 \tan{\left( \theta_0 -\theta'_1 \right)} - R \left( \sin{\theta_0} - \sin({2\theta_2 - \theta_1}) \right) 
    \\
    &-L_2 \tan{2\left(\theta_2-\theta_1 \right)}.
\end{split}   
    \label{eq:vray}
\end{equation}
%-----------------------------
In the magnified view, the line segment $P_{\varepsilon}P_2$ intersects the line segment $P_3P_4$ at a right angle.
Applying the Pythagorean theorem to the right triangle $P_3P_4P_2$, we obtain
%-------------------------------
\begin{equation}
\begin{split}
    &{P_3P_2}^2 
    = P_3P_4^2 + P_4P_2^2
\\
&\left( 2R_{\rm drop} \sin{\left( \frac{\theta_0-(2\theta_2-\theta_1)}{2} \right)} \right)^2 = 
 \left( \frac{(L_3 + l_b)\sin{\varepsilon_y} }{\cos{(\varepsilon_y + \theta_2-\theta_1})} 
   \right)^2 + 
 \left( \frac{L_2 + l_b}{\cos{(\theta_2-\theta_1})}  - \frac{(L_3 + l_b) \cos{\varepsilon_y}}{\cos{(\varepsilon_y + \theta_2-\theta_1})} \right)^2.
\end{split}   
    \label{eq:P2P3P4}
\end{equation}
%----------------------------

The deviation angle at point $P_{\varepsilon}$ is estimated by solving the non-linear equations (Eqs.~\eqref{eq:snell}–\eqref{eq:P2P3P4}).
By sampling points of $P_{\varepsilon}$ along the $x$-axis over the range $-R_{\rm drop}<x<R_{\rm drop}$, the $y$-deviation angle on the $x$-$y$ plane can be obtained for the entire range.
Similarly, on the $x$-$z$ plane, the deviation angle in the $x$-direction can be estimated at each sampled point using the same procedure as in Eqs.~\eqref{eq:snell}--\eqref{eq:P2P3P4}.

%-------------------------------------------
\begin{figure}[!ht]
    \centering
    \includegraphics[width=0.5\linewidth]{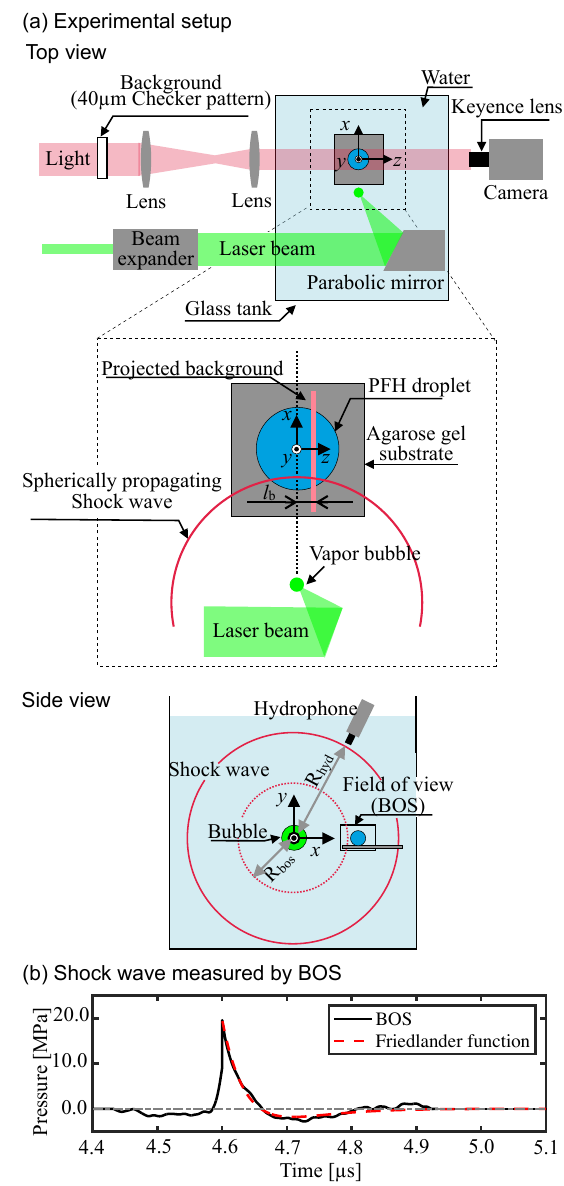}
    \caption{(a) Experimental setup of the BOS technique. 
    Top view: The center of the droplet is defined as the origin $O$. 
    A light source, a background with a checker pattern, two lenses, the droplet, and a camera are aligned along the $z$-axis. 
    The background pattern is projected into the droplet through the two lenses. 
    An enlarged view of the droplet region is shown within the dashed rectangle. 
    The droplet is placed on an agarose gel substrate immersed in water. 
    A vapor bubble and a spherically propagating shock wave are generated by a laser focused along the $x$-axis. 
    Side view: A hydrophone is positioned at $R_{\rm hyd}$, which was a safe distance from the shock source to record the shock wave.
   (b) The black line indicates shock pressure at $R_{\rm bos}$ measured by the BOS in the absence of both the droplet and the agarose gel substrate.
    The dashed red line represents the pressure profile estimated using the Friedlander function (Eq.~\eqref{eq:friedlanderEquation}), based on the maximum pressure value of the black line. The gray dashed line with a single dot indicates 0~MPa.}
    \label{fig:exp}
\end{figure}
%-------------------------------------------

%%===============================================
\section{Methods}\label{sec:exp} 
\subsection{Experimental setup}\label{sec:exp-setup}

\begin{table*}[t]
    \centering
    \caption{The constant parameters required to define the NASG--EoS and Tait's equation for PFH and water.}
    \begin{tabular*}{0.6\textwidth}{@{\extracolsep{\fill}}lcc}
        \specialrule{1.2pt}{0pt}{0pt}
         & PFH & water \\
        \midrule
        $n$ [-] & 1.25 & 1.33 \\ 
        $K$ [m$^3$ kg$^{-1}$] \cite{gladstone1863xiv} & $1.49 \times 10^{-4}$ & $3.34 \times 10^{-4}$ \\
        $c$ [m s$^{-1}$] & 509.8 & 1500 \\
        $\gamma$ [-] & 1.477 & -- \\
        $\kappa$ [-] & -- & 7.15 \\
        $b$ [-] & $2.042 \times 10^{-4}$ & -- \\
        $C_v$ [J /$(\rm kg \cdot K)$] & 515.615 & -- \\
        $P_\infty$ [Pa] & $1.7561 \times 10^{8}$ & $314 \times 10^{6}$ \\
        $p_0$ [Pa] & 101325 & 101325 \\
        $\rho_0$ [kg m$^{-3}$] & 1680.1 & 1006.21 \\
        \specialrule{1.2pt}{0pt}{0pt}
    \end{tabular*}
    \label{tb:parameter}
\end{table*}

 %-------------------------------------------
The experimental setup is illustrated in Fig.~\ref{fig:exp}.  
A PFH droplet with a radius of around 1 mm was placed on an agarose gel substrate with a width of several tens of millimeters, immersed within a tank filled with distilled water.  
The constant parameters of PFH and water are listed in Table~\ref{tb:parameter}.
A pulsed Nd:YAG laser (Lumibird Quantel Q-smart, 6 ns, 532 nm) was expanded using a 10~$\times$ beam expander (52-71-10X-532/1064, Special Optics) and focused in the water by a  90$^{\circ}$ parabolic mirror (Edmund Optics, $f$~=~50~mm) to induce a plasma that results in the formation of a vapor bubble and spherically propagating shock wave.  
The laser focus and the PFH droplet were positioned to be aligned along the $x$-axis, where the origin of the $xyz$-coordinate was set at the center of the droplet.

The components enabling the BOS measurement included a light source (CAVILUX Smart, HF, Nobby Tech. Ltd.), a background printed with a 40 µm checker pattern, two achromatic lenses (63701 and 32925, Edmund Optics Co. Ltd.), a camera lens (VH-Z50L, KEYENCE Ltd.), and a high-resolution, single-frame camera (Nikon D7200, spatial resolution : 4000 $\times$ 6000 pixels), all aligned along the $z$-axis.  
The background was projected inside the droplet at $z = l_b$ in the $x$-$z$ cross-section using two achromatic lenses.
Here, $l_b$ was 0.16 mm and the distance between these two lenses was around 300 mm.
The camera’s focal plane is aligned with the image plane of the projected background.
The camera captures both the reference and distorted images, as shown in Fig.~\ref{fig:exp-displacement}(a).  
The resolution of image is 0.45 µm/pixel.

% ----------------------------------------
\begin{table*}[t] 
    \centering 
    \caption{Arrival times $t_{\rm hyd}$ and $t_{\rm bos}$, and peak pressures $p_{\rm hyd}$ and $p_{\rm bos}$, measured by the hydrophone and BOS. 
    The peak pressure of the shock wave at $t_{\rm bos}$, $p_{\rm e,bos}$, estimated from the hydrophone measurements using Eq.~\eqref{eq:hyd}, is also shown.}
    
    \begin{tabular*}{\textwidth}{@{\extracolsep{\fill}}ccccc}
    \specialrule{1.2pt}{0pt}{0pt} 
    $t_{\rm hyd}$ [µs] & $t_{\rm bos}$ [µs] & $p_{\rm hyd}$ [MPa] & $p_{\rm e,bos}$ [MPa] & $p_{\rm bos}$ [MPa] \\ 
    \midrule 
    17.2 $\pm$ 1.4 & 4.60 $\pm$ 0.01 & 2.2 $\pm$ 0.7 & 15.0 $\pm$ 4.8 & 19.6 $\pm$ 0.5 \\ 
    \specialrule{1.2pt}{0pt}{0pt} 
    \end{tabular*}    
    \label{tab:rest_hyd_bos}
\end{table*}

% ----------------------------------------

Here, we employed a synchronization technique, the details of which are described in Ichihara \textit{et al.} \cite{ichihara2025high}, to capture the high-speed phenomena.
A delay generator (Stanford Research Systems, DG645) was used to synchronize the timings between the camera, the light source, and the laser.
As the delay generator received an external trigger from laser irradiation, it sent trigger signals to the camera and light source.
The camera shutter opened simultaneously with laser irradiation, while the light source illuminated with $t$ of delay after the laser irradiation.
The illuminating light was emitted for 10 ns, and the delay time $t$ was varied from 4.5 to 11.0~µs to capture different temporal stages of the event.
It should be noted that three different droplet radii were employed in this experiment to enable high-speed measurements.
The droplet radii $R_{\rm drop}$ were 0.91, 1.00, and 1.02 mm, and the corresponding droplet ellipticities $e$ were 0.04, 0.04, and 0.08, respectively.
Given that the ellipticities $e$ were close to zero, the droplets were assumed to be spherical in this experiment.

A hydrophone was used to estimate the maximum pressure of the shock wave in the absence of the droplet within the field of view in the BOS setup.
To avoid damage from the shock wave, the hydrophone was positioned at a larger distance away from the shock wave source, as shown in Fig.~\ref{fig:exp} (side view) \cite{supponen2017shock}.
Let $R_{\rm hyd}$ and $R_{\rm bos}$ be the distances from the vapor bubble to the hydrophone and to the field of view (BOS), respectively.
The hydrophone was synchronized with the timing of laser irradiation using a delay generator, and it measured the arrival time $t_{\rm hyd}$ and the maximum pressure $p_{\rm hyd}$.

These measurements were repeated 10 times under identical conditions, and the averaged values were used for further analysis.
The peak pressure of a shock wave propagating in water reduces in distance, due to nonlinear dissipation caused by inelastic heating and gaseous contaminants, in addition to spherical spreading.  
Based on the results of Vogel \textit{et al.} (1996) and Bokman \textit{et al.} (2023), the decay of the peak pressure with respect to distance in the far field is given by
\cite{vogel1996shock,bokman_scaling_2023}:
\begin{equation}
    p_{\rm e,bos} = p_{\rm hyd}\left( \frac{R_{\rm bos}}{R_{\rm hyd}} \right)^{-1.4},
    \label{eq:hyd}
\end{equation}
where the distances $R_{\rm bos}$ and $R_{\rm hyd}$  are estimated using the sound speed in water and the corresponding arrival time, $t_{\rm bos}$ and $t_{\rm hyd}$, respectively.

BOS measurements were conducted to independently validate the pressure obtained by hydrophone.
Fig.~\ref{fig:exp}(b) shows the pressure recorded by BOS at $t_{\rm bos} =$~4.6~µs. 
Table \ref{tab:rest_hyd_bos} summarizes the peak pressures and arrival times obtained from the hydrophone and BOS measurements.
In this measurement, the curvature of the shock front was negligible due to the narrow field of view.
%, and the front propagated essentially normal to the field of view. 
Therefore, the density gradient field was obtained by dividing the displacement field by the shock wave thickness of $2\Delta l_b$ (=0.12~mm).
At 4.6~µs, the shock front exhibited a discontinuity at its peak in the density gradient field. 
Accordingly, the density was reconstructed by splitting the gradient at its maximum value and integrating from the pre-shock and post-shock sides separately. 
The density field, both before 4.4~µs and after 4.9~µs, was set equal to the density of the ambient fluid $\rho_0$ because these regimes correspond to a quasi-steady field.
To compute the pressure from the reconstructed density, Tait’s equation was used:
\begin{align}
    \frac{p + P_{\infty}}{p_0 + P_{\infty}} = \left( \frac{\rho}{\rho_0} \right)^{\kappa},
\end{align}
where $p_0$ is the atmospheric pressure, and $P_{\infty}$ and $\kappa$ are constants for standard-state water \cite{brujan2010cavitation}. 
The overall procedure for reconstructing the pressure follows Yamamoto \textit {et al.}~\cite{yamamoto2022contactless}.

As shown in Table~\ref{tab:rest_hyd_bos}, the maximum pressure estimated from the hydrophone measurements is in good agreement with the maximum pressure obtained by BOS. Therefore, the initial shock-wave pressure used in the numerical simulations was set to the value determined by the BOS measurements.

\subsection{Data analysis procedure}\label{sec:exp-analysis}

\begin{figure}[ht]
    \centering
    \includegraphics[width=0.5\linewidth]{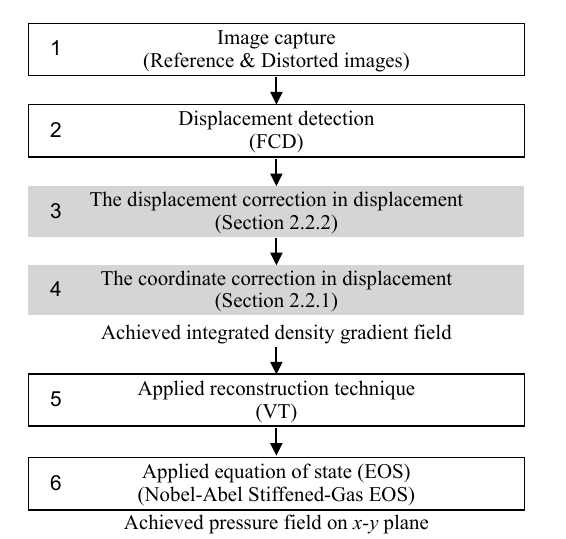}
    \caption{The steps of the data analysis procedure of the BOS technique, from the captured images to the reconstructed pressure field.}
    \label{fig:exp-analysis_step}
\end{figure}

To estimate the pressure distribution from the captured images, image processing is applied to both the reference and distorted images.
The overall procedure to calculate pressure from the detected displacement is outlined in Fig.~\ref{fig:exp-analysis_step}.
Section \ref{sec:calculation_step-disp} describes the displacement detection and the correction of optical distortion caused by the refractive index mismatch between the droplet and the surrounding water (Fig.~\ref{fig:exp-analysis_step} steps 1--4).
Section \ref{sec:calculation_step-pre} presents the method for converting the corrected displacement into pressure (Fig.~\ref{fig:exp-analysis_step} steps 5--6).
Section \ref{sec:calculation_uncertainty} describes the uncertainty arising from error propagation in the data analysis procedure.

% ----------------------------------------
%-=-=-=-=-=-=-=-=-=-=-=-=-=-=-=-=-=-=-=-=-=-=-=-=
\subsubsection{Displacement detection}\label{sec:calculation_step-disp} 

\begin{figure}
    \centering
    \includegraphics[width=0.5\linewidth]{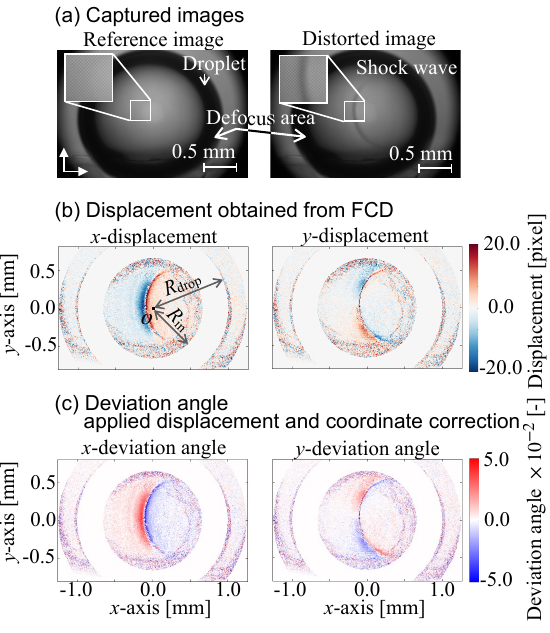}
    \caption{(a)~Captured droplet images during the experiment.
    The left image corresponds to the case in the absence of a shock wave, while the right image shows a droplet in the presence of a shock wave.
    (b)~Displacement fields calculated from the images in~(a). 
    The left and right panels represent the non-corrected displacements in the $x$- and $y$-directions, respectively. 
    $R_{\rm drop}$ denotes the droplet radius, and $R_{\rm in}$ indicates the inner radius of the droplet.
    The displacement values in the region between $R_{\rm in}$ and $R_{\rm drop}$ were set to zero, as measurements in this zone are unreliable due to strong defocus effects.
    (c)~Deviation angle calculated from the displacement field~(b), with displacement and coordinate corrections applied to account for droplet-induced distortion and spatial misalignment of the coordinate system.}
    \label{fig:exp-displacement}
\end{figure}

The displacement field $\boldsymbol{w_{\rm ray}} = (u_{\rm ray}, v_{\rm ray})$ was obtained from the captured images (Fig.~\ref{fig:exp-displacement}(a)) using the Fast Checker Demodulation (FCD) method \cite{wildeman2018real, shimazaki2022background}, which enables high-resolution measurement of large displacements by using a systematic pattern and Fast Fourier transform, as shown in Fig.~\ref{fig:exp-displacement}(b). 
The measurable maximum displacement is 21.2~pixel in this experimental conditions.
Herein, the displacement $\boldsymbol{w_{\rm ray}}$ was set to zero in the region between the inner radius $R_{\rm in}$ and the droplet radius $R_{\rm drop}$ to exclude regions where the displacement measurement is unreliable due to strong defocus, where $R_{\rm in} = 0.66 R_{\rm drop}$. 

This displacement field $\boldsymbol{w_{\rm ray}}$ contains distortions caused by the shock wave as well as by the droplet itself, due to the difference in refractive index between the droplet and the surrounding fluid.  
To remove the droplet-induced distortion, a displacement correction was applied using Eqs.~\eqref{eq:snell}--\eqref{eq:P2P3P4}.  
The constants and geometric parameters used in these equations are provided in Section~\ref{sec:exp-setup}.  
The droplet radius $R_{\rm drop}$ was defined as the average of its semi-major and semi-minor axes, i.e., $R_{\rm drop} = (r_s + r_l)/2$, where $r_s$ and $r_l$ denote the semi-minor and semi-major axes, respectively. 
These parameters were estimated from the reference image using the MATLAB function \textit{fit\_ellipse}.

The deviation angle obtained after this correction still contains a component associated with the spatial misalignment of the coordinate system caused by the refractive effect of the curved droplet interface.  
To account for this, a coordinate correction (Eq.~\eqref{eq:ray-PoPb}) was applied to remove the spatial misalignment.  
After these corrections for both distortion and coordinate misalignment, the deviation angle $\boldsymbol{\varepsilon}$ (Fig.~\ref{fig:exp-displacement}(c)) satisfies the conventional BOS theory (Eq.~\eqref{eq:bos_epsilon}).
% ----------------------------------------
% ----------------------------------------

%----------------------------------------------------
\begin{figure}
    \centering
    \includegraphics[width=0.5\linewidth]{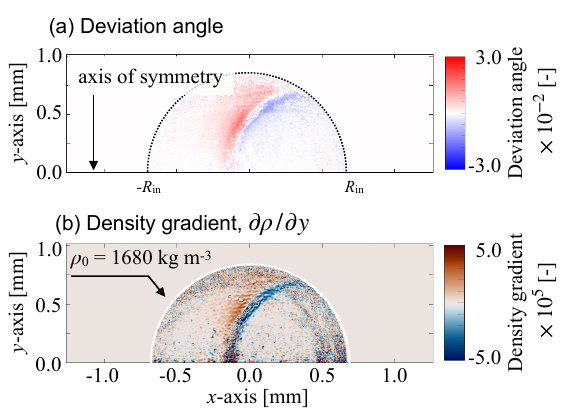}
    \caption{Method for deriving pressure from deviation angle $\varepsilon_y$.
    (a) The deviation angle field $\varepsilon_y$ is symmetrically divided at the axis $y=0$ and averaged over the regions $y<0$ and $y>0$. 
    The region outside the effective radius $R_{\rm in}$ is assumed to exhibit zero displacement.
    (b) Density gradient field computed by applying VT to the deviation angle field in (a).
    The white semicircle indicates the boundary used when integrating the density gradient field to obtain the density, where the boundary condition is set to the PFH density $\rho_0 = $1680.1 kg/m$^3$.}
    \label{fig:exp-analysis_condition}
\end{figure}
%----------------------------------------------------
%-=-=-=-=-=-=-=-=-=-=-=-=-=-=-=-=-=-=-=-=-=-=-=-=
\subsubsection{Pressure computation}\label{sec:calculation_step-pre}
Vector tomography (VT) was employed to reconstruct the density gradient field on the $x$-$y$ plane \cite{ichihara2025high}.
The axis of symmetry required for VT was defined along the $x$-axis crossing the  $O$.
Therefore $y$-direction deviation angle $\varepsilon_y$ was used in VT.
The deviation angle $\varepsilon_y$ field was divided about this axis of symmetry. 
One side was reflected about the axis and then averaged with the other side to obtain the mean field, as shown in Fig.~\ref{fig:exp-analysis_condition}(a).
This averaged deviation angle was used for the VT.
The following matrix formulation expresses the relationship between the deviation angles and the density gradients at a given $x$-axis,
% -------------------------
\begin{equation}
%\scalebox{0.7}{$
\begin{pmatrix}
\varepsilon^{1}_{y} \pm s \\
\varepsilon^{2}_{y} \pm s\\
\vdots \\
\varepsilon^{N-1}_{y} \pm s\\
\varepsilon^{N}_{y} \pm s\\
\end{pmatrix}
=
\begin{pmatrix}
\alpha_{1,1} &\alpha_{1,2} &\alpha_{1,3} & \cdots& \alpha_{1,N} \\
0 &\alpha_{2,2} & \alpha_{2,3}& \cdots& \alpha_{2,N}\\
\vdots&\vdots&\ddots&\ddots&\vdots \\
0 &0 &  \cdots&\alpha_{N-1,N-1}& \alpha_{N-1,N}\\
0&0&\cdots&0&\alpha_{N,N}\\
\end{pmatrix}
\begin{pmatrix}
\rho'_1 \pm s^{1}_{\rho'}\\
\rho'_2 \pm s^{2}_{\rho'}\\
\vdots \\
\rho'_{N-1} \pm s^{N-1}_{\rho'}\\
\rho'_N \pm s^{N}_{\rho'}\\
\end{pmatrix}
%$}.
\label{eq:vt-bos}
\end{equation}
% -------------------------
Here, the superscripts denote the $y$-coordinate, while $N$ indicates the largest $y$-coordinate from the symmetry axis. 
$s^{k}_{\rho'}$ represents uncertainty propagated from in deviation angle, $k$ is a natural number ranging from 1 to $N$, and $s$ denotes the uncertainty in the deviation angle along the $y$-direction, the details of which are provided in Section~\ref{sec:calculation_uncertainty}.
The constant matrix $\boldsymbol{\alpha}$ depends on $N$.
Using Eq.~\eqref{eq:vt-bos}, the inverse of the $\boldsymbol{\alpha}$ matrix is solved to obtain the density gradient from the deviation angle at a given $x$-axis.
Furthermore, by applying Eq.~\eqref{eq:vt-bos} to the deviation angles along the $x$-axis in the range -1.5~mm~$< x < 1.5 \ \rm{mm}$, the distribution of the density gradient $\rho' (= \partial \rho/ \partial y)$ on the $x$-$y$ plane passing through the origin (Fig.~\ref{fig:exp-analysis_condition}(b)) is obtained.
The density field was then calculated from the density gradient using the trapezoidal rule.
As a boundary condition for this integration, the density at the inner radius $R_{\rm in}$ was set equal to the reference PFH density $\rho_0$.

Finally, the pressure field was obtained from the reconstructed density field using the NASG--EoS Eq.~\eqref{eq:bos_nasg-eos}.
The overall procedure from displacement to pressure reconstruction presented in this section follows Ichihara \textit{et al.} \cite{ichihara2025high, ichihara2022background}.

% ---------------------------------------
\subsection{Uncertainty in BOS}\label{sec:calculation_uncertainty}
\begin{figure}[ht]
    \centering
    \includegraphics[width=0.5\linewidth]{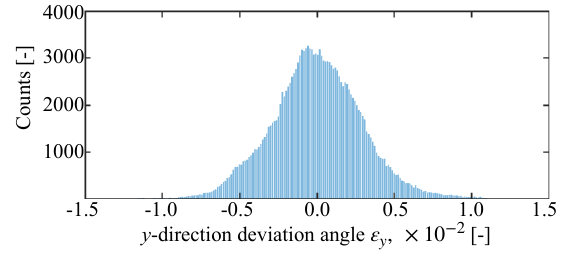}
    \caption{Histogram of the deviation angle in the $y$-direction for single dataset.
    The deviation angle was obtained from two regions of 301 $\times$ 281
    pixels extracted on the left and right sides outside the droplet radius $R_{\rm drop}$.
    These regions can be regarded as representing a quasi-steady field and are unaffected by the shock wave.}
    \label{fig:exp-uncertainty}
\end{figure}

This section presents the procedure and results of the uncertainty analysis for the pressure field derived from the deviation angle.
Initially, two regions of $301 \times 281$ pixels were extracted from the left and right sides of the deviation angle field $\varepsilon_y$, outside the droplet radius $R_{\rm drop}$. 
Fig.~\ref{fig:exp-uncertainty} shows the histogram of the $y$-direction components of the deviation angle within this sample regions. 
These sample regions correspond to a quasi-steady field and occupy 0.84~\% of the entire field of 5601 $\times$ 3601 pixels.
In principle, the deviation angle should be zero in such a quasi-steady field.

As observed in Fig.~\ref{fig:exp-uncertainty}, the histogram follows a Gaussian distribution, indicating that the deviation angle measurements exhibit statistically consistent behavior. 
The measured deviation angles can be considered free from systematic errors and are dominated by random errors. 
The corresponding uncertainties $s$, estimated from the $y$-direction deviation angle, are $2.9 \times 10^{-3}$ and $5.9 \times 10^{-4}$ for the single dataset and for the dataset averaged over ten measurements, respectively.
Here, the uncertainty $s$ indicates that the deviation angle has a 68 \% probability of falling within $s$ on either side of zero, $\varepsilon_y \pm s$.
The discretization error introduced by the displacement/coordinate correction procedure described in Section~\ref{sec:calculation_step-disp} is sufficiently small compared with these uncertainties and is therefore neglected.
%----------------------------------------------------
%----------------------------------------------------

Next, the procedure used to evaluate the propagation error —from the deviation angle to the pressure field — is described.
The deviation angle is first subjected to vector tomography (VT), from which the density gradient is obtained using Eq.~\eqref{eq:vt-bos}.
In the VT matrix, the density gradient is computed by solving the inverse of an upper triangular matrix.
The gradient components are evaluated sequentially from $y=n$ toward $y=1$, corresponding to $\rho'_{n},\rho'_{n-1},\rho'_{n-2}, \cdots$.
The uncertainties do not propagate linearly through this forward substitution, and the accumulated uncertainty increases toward the axis of symmetry.
Taking this behavior into account, Eq.~\eqref{eq:uncertainty-vt} defines the uncertainty $s^{k}_{\rho'}$ in density gradient at a given position $y=k$:
% ----------------------------
\begin{equation}
s^{k}_{\rho'} = \pm 
\frac{s}{\alpha_{k,k}}
\sqrt{  1+ \sum^{n-1}_{j=k}  \left( \frac{\alpha_{j,j+1} s^{k+1}_{\rho'}}{s} \right)^2  }.
\label{eq:uncertainty-vt}
\end{equation}
% --------------------------------
When the contribution $\left( \alpha_{j,j+1} s^{k+1}_{\rho'} \right)^2$ is smaller than 0.01, it is regarded as negligible and set to zero.

The density field was obtained by integrating the density gradient using the trapezoidal rule.
The uncertainty in the density $\rho_k$ at a given position $y=k$, denoted as $s^{k}_{\rho}$, is expressed as
% ----------------------------
\begin{equation}
s^{k}_{\rho}=
\pm 
dx \sqrt{
\sum^{k}_{j=1} 
A^2_j}.
\label{eq:uncertainty-integration}
\end{equation}
% --------------------------------
Here, $dx$ denotes the resolution of image.
% $k$ is a natural number ranging from 1 to $N$, and 
The approximation error of the trapezoidal rule scales with the square of the spatial resolution.
Its magnitude is $O(10^{-2})$ smaller than that of the uncertainty $s^{k}_{\rho}$; hence, its contribution is neglected.

The uncertainty in the pressure obtained using the NASG--EoS, denoted as $s^{k}_{p}$, can be evaluated from
% ----------------------------
\begin{equation}
s^{k}_{p} = 
\pm (\gamma-1)C_v T_0\frac{s^k_{\rho}}{(1-\rho_k b)^2},
\label{eq:uncertainty-eos}
\end{equation}
% --------------------------------
where $\rho_k$ represents the density at $y=k$.
In this analysis, both $s^{k}_{\rho}$ and $s^{k}_{p}$ were evaluated under the assumption that the uncertainties $s$ constitute mutually independent random errors.

%----------------------------------------------------
\begin{figure}[ht]
    \centering
    \includegraphics[width=0.5\linewidth]{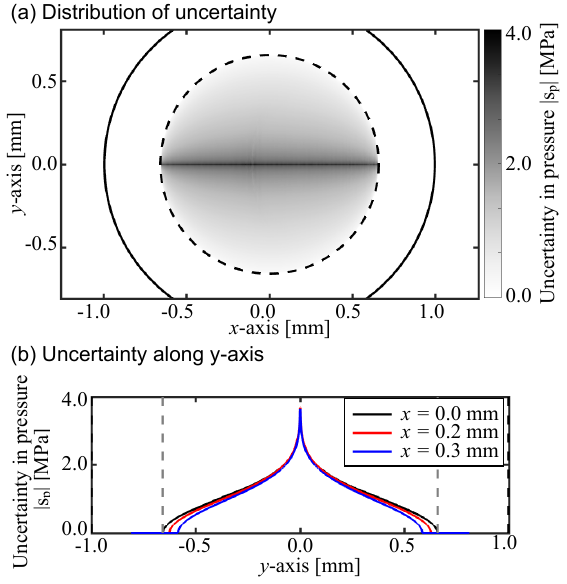}
    \caption{Uncertainty in the pressure field $|s_p|$ obtained from the deviation angles for single dataset, where absolute values are shown.
    (a) Distribution of pressure uncertainty; (b) uncertainty profiles along the $y$-axis at $x = $0, 0.2, 0.3 mm.}
    \label{fig:exp-uncertainty_pressure}
\end{figure}
%----------------------------------------------------
The pressure uncertainty $|s_p|$, calculated from the uncertainty $s$ in the deviation angle for single dataset, is presented in Fig.~\ref{fig:exp-uncertainty_pressure}. 
$|s_p|$ denotes the absolute magnitude of the pressure uncertainty. 
Fig.~\ref{fig:exp-uncertainty_pressure}(a) shows the spatial distribution of $|s_p|$, whereas Fig.~\ref{fig:exp-uncertainty_pressure}(b) indicates the uncertainty along $-1.0 < y < 1.0$ mm at $x = $ 0, 0.2, and 0.3 mm. 
As seen in Fig.~\ref{fig:exp-uncertainty_pressure}(a), the uniformly distributed uncertainty $s$ in the deviation angle field increases near the axis of symmetry due to the inverse matrix $\boldsymbol{\alpha}^{-1}$ in the VT. 
Fig.~\ref{fig:exp-uncertainty_pressure}(b) further indicates that $|s_p|$ reaches approximately 3.5~MPa near the axis of symmetry, while it remains below 10~MPa in the vicinity of the inner radius $R_{\rm in}$. 
In addition, for a given $y$-coordinate, the uncertainty $|s_p|$ slightly decreases as the position moves away from the origin along the $x$-direction. 
The distribution of pressure uncertainty is unaffected by whether averaging is applied; however, its magnitude scales proportionally with the uncertainty in the deviation angle. 
For the pressure calculated from the averaged deviation angles, the uncertainty on the axis of symmetry ($y$=0) was approximately $\pm 0.3$~MPa.

%------------------------------------

\subsection{Simulation}\label{sec:exp-simulation}

 Numerical simulations of the shock wave-droplet interaction have been carried out to validate the experimental results by comparing the measured density gradient field with the BOS measurements. The multi-phase, diffuse-interface-method-based Euler equation solver implemented in ECOGEN \cite{schmidmayer_ecogen_2020} has been selected to correctly model shock wave focusing in the droplet. The model is detailed in Kapila~\textit{et al.} \cite{Kapila2001ECOGENModel} and Saurel~\textit{et al.} \cite{Saurel2009ECOGENModel} and has been previously validated for several multiphase problems \cite{pishchalnikov_high-speed_2019, dorschner_formation_2020,fiorini2026cavitation}.
% Herein, this numerical simulation follow to Fiorini \textit{et al.} \cite{fiorini2025positive}

\begin{figure}[ht]
    \centering
\includegraphics[width=0.5\linewidth]{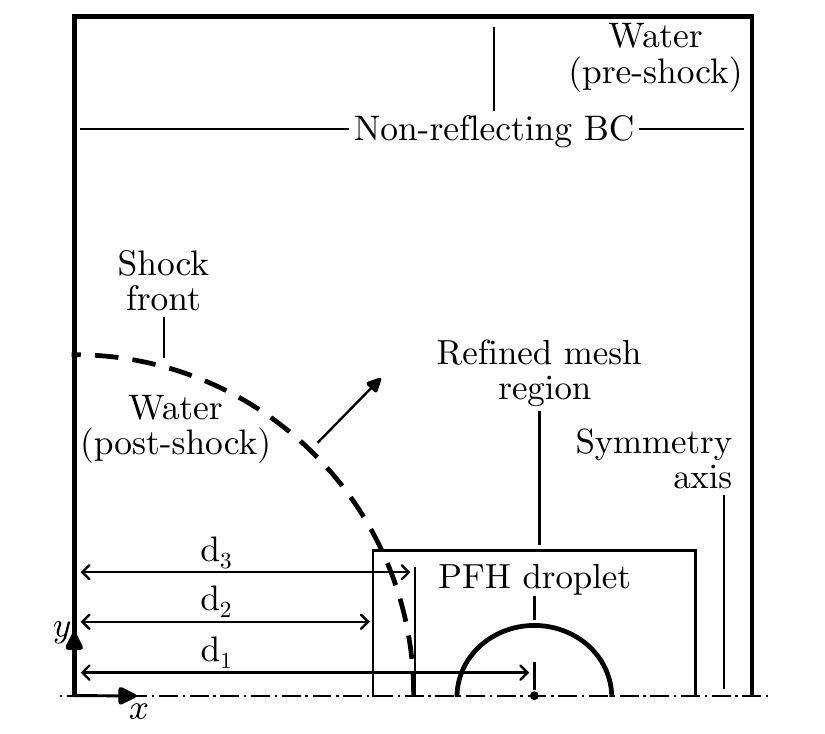}
    \caption{Schematic of the numerical domain and initial conditions. An ellipsoidal PFH droplet lies on the symmetry axis at the domain's lower edge. The droplet's center is at an horizontal distance $ d_1 $ from the origin of the domain, which is located at its bottom-left corner. The post-shock state is defined in Eq.~(\ref{eq:initial_pressure}) such that the wave front is at a distance $ d_2 $ from the origin and is traveling from left to right. Non-reflecting boundary conditions are applied on the left, right and top edge. \\ }
    \label{fig:num-simulation_setup}
\end{figure}

The simulation domain, designed to reproduce the experimental configuration, is presented in Fig.~\ref{fig:num-simulation_setup} and consists in a rectangle of size 14$\times$14 \si{\milli\meter}. 
The numerical domain has been approximated as two-dimensional axisymmetric with respect to the $x$-axis (bottom edge). 
Non-reflective boundary conditions are enforced on the left, top, and right side. 
The PFH droplet is modeled as an ellipsoid (an ellipse when projected onto the two-dimensional domain) to account for the slight deformation due to gravity. 
The semi-major axis is positioned along the bottom edge of the domain and has a value $ r_l = 509\  \si{\micro\meter} $, while the semi-minor axis points to the $y$-direction and has a magnitude $ r_s = 488.25\  \si{\micro\meter} $. 
The center of the droplet is located at a distance $ d_1 = r_l + 7.050\ \si{\milli\meter} $ from the left side of the domain. 
A rectangular $ 4r_l \times 2r_s$ box with cell resolution $ \Delta x = \Delta y = 5\ \si{\micro\meter} $ is initialized around the droplet, with its bottom edge lying on the $x$-axis, and the left edge located at a distance $ d_2 = d_1 - r_l $. 
 Mesh stretching with a geometric progression of 1.05 is applied outside of that box to optimize computational time. Adaptive mesh refinement is also employed, with a maximum of 4 refinement levels \cite{schmidmayer2019adaptive}.

Thermodynamic properties of water are modeled using the Stiffened-Gas Equation of State (SG EoS) as described in Le Métanier~\textit{et al.}~\cite{le_metayer_elaboration_2004}, while the NASG--EoS is employed to model PFH properties. 
The values of the thermodynamic constants, as well as the procedure followed to fit the parameters for the PFH equation of state are available in the work by Prasanna~\textit{et al.}~\cite{Prasanna2025PFHEquationOfState}. 
The initial particle velocities of both water and PFH, $u_{x,0}$ and $u_{y,0}$, are set to zero, assuming a uniform temperature of $T = 298\ \si{\kelvin}$.
The remaining initial conditions are listed in Table~\ref{tb:parameter}.

\begin{table*}[ht]
    \centering
    \caption{Parameters defining the initial condition for the water's shocked state in the numerical domain.}
    \begin{tabular}{ccccc}
        \specialrule{1.2pt}{0pt}{0pt}
         $ x_I $ [\si{\meter}] & $ \beta $ [-] & $ p_{\mathrm{bos}} $ [\si{\pascal}] &$\rho_{\mathrm{max}} [\si{\kilo\gram\per\meter\cubed}]$ &  $u_{x, \mathrm{max}}\ [\si{\meter\per\second}]$ \\
        \midrule
        88.23$\times10^{-6}$ & 1.177 & $ 19.57\times10^6 $ &1015.31 &  13.168\\        
        \specialrule{1.2pt}{0pt}{0pt}
    \end{tabular}
    \label{tab:initial_condition_parameter}
\end{table*} 
%---------------------

The radially propagating shock wave front has a radius of curvature of $ 7\ \si{\milli\meter} $ and is initialized at a distance $ d_3 = 7\ \si{\milli\meter} $ from the left edge of the domain. 
The pressure profile obtained from the BOS measurements on the $x$-axis is least-square fitted to a Friedlander function:
\begin{equation}
        p(x) = p_{\mathrm{bos}}(1 - x/x_I) e^{-\beta x/x_I},
        \label{eq:friedlanderEquation}
        \end{equation}
where the target width, $x_I$, is determined from the distance from the shock front at which the pressure profile crosses the zero-pressure line based on the BOS measurements.
The parameter $ \beta $ characterizes the exponential decay of the pressure signal.
The resulting values are reported in Table~\ref{tab:initial_condition_parameter}. 
The pressure profile estimated using the Friedlander function (Eq.~\eqref{eq:friedlanderEquation}) is shown as the dashed red line in Fig.~\ref{fig:exp}(b).

Once the parameters are determined, the pressure profile is initialized in the simulation domain using the function:
 \begin{equation}
        p(r) = (p_{\mathrm{bos}} - p_0) \left(1 - \frac{d_3 - r}{x_I} \right)e^{-\beta (d_3 - r)/x_I} + p_0
        \label{eq:initial_pressure}
    \end{equation}
assuming radial symmetry of the shock front, with $ r = x^2 + y^2 $ evaluated on the interval $ [0, d_3]$ and the origin located at the bottom-left corner of the domain. The pressure $ p_0 $ is added to account for the atmospheric pressure.

The initial condition for the density and the velocity mixture can be obtained in an analogous way:
\begin{equation}
    \rho(r) = (\rho_{\mathrm{max}} - \rho_{0}) \left(1 - \frac{d_3 - r}{x_I} \right)e^{-\beta (d_3 - r)/x_I} + \rho_{0}, \\
    \label{eq:initial_density}
\end{equation}
\begin{equation}
    u_{r}(r) = (u_{x, \mathrm{max}} - u_{x, 0}) \left(1 - \frac{d_3 - r}{x_I} \right)e^{-\beta(d_3 - r)/x_I} + u_{r, 0}, \\
    \label{eq:initial_speed}
\end{equation}
where the values for the peak density $ \rho_{\mathrm{max}} $ and peak horizontal velocity $ u_{x, \mathrm{max}} $ are calculated using the Rankine-Hugoniot jump conditions. The values can be found in Table~\ref{tab:initial_condition_parameter}.

%%===============================================
\section{Results and Discussion}\label{sec:resutl}
%---------------------------------------
\subsection{Spatiotemporal integrated density gradient}\label{sec:result-densitygradient}
\subsubsection{Integrated density gradient field}\label{sec:result-densitygradient_field}

\begin{figure}[ht]
    \centering
    \includegraphics[width=0.9\linewidth]{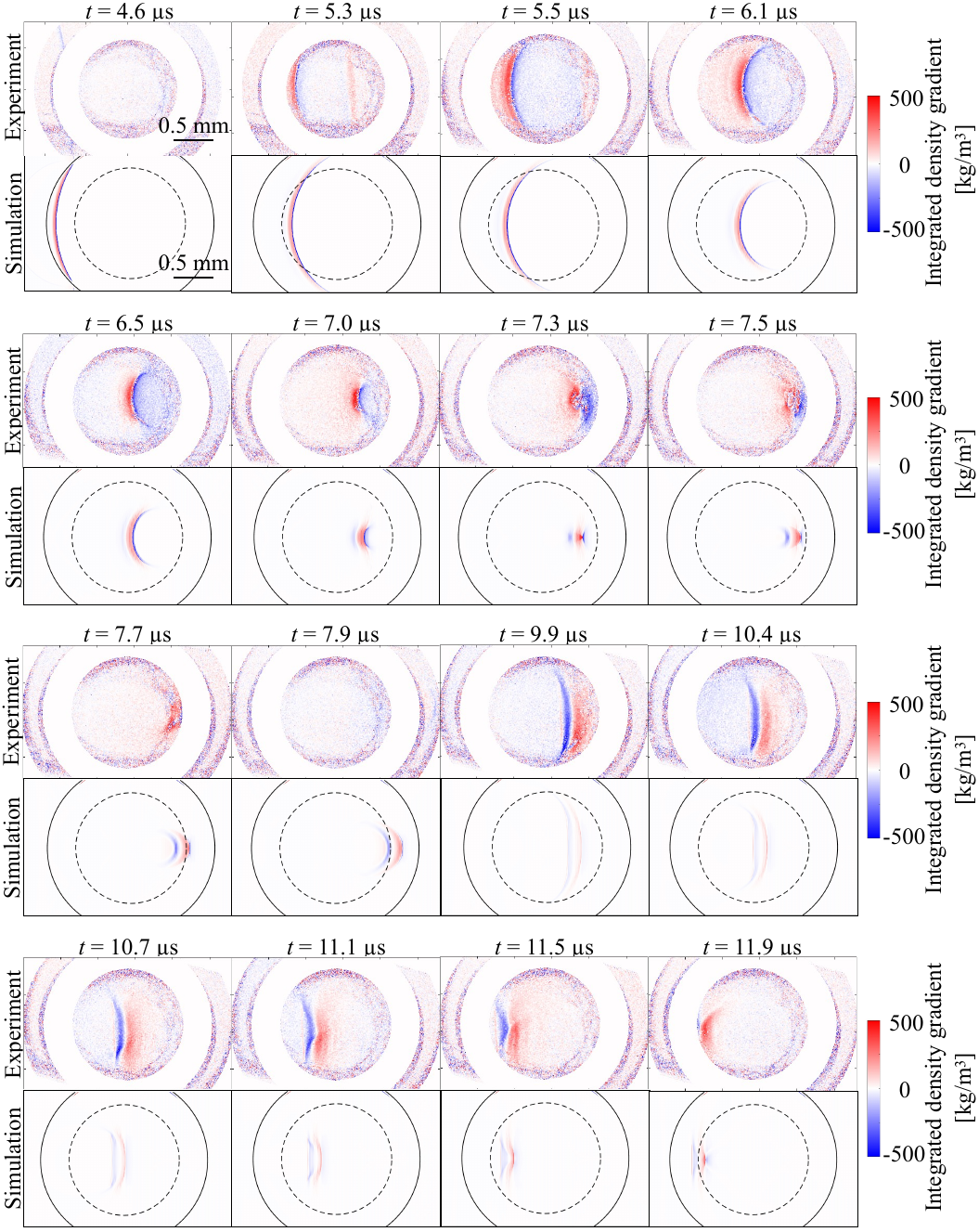}
    \caption{The $x$-direction density gradient, integrated along the $z$-axis.
    The horizontal and vertical axes represent $x$ and $y$, respectively.
    The top panels show the fields measured by BOS, while the bottom panels present those obtained from ECOGEN simulations. 
    The instant of shock wave generation defines $t$ = 0. Between $t$ = 4.6~µs and 7.9~µs, the shock wave propagates from left to right. For $t$~=~9.9~µs, the shock wave reflected  at the PFH–water interface propagates from right to left.}
    \label{fig:result-density-field}
\end{figure}

The $x$-direction density gradient, integrated along the $z$-axis, is shown in Fig.~\ref{fig:result-density-field} for different time instants. 
The top panels show the measurement results of BOS, while the bottom panels present numerical simulations of a shock wave propagating through the droplet from $t$ = 4.6--11.9~µs. 
For the BOS results, the integrated density field was obtained from the corrected integrated deviation angle $\varepsilon_x$ using Eqs.~\eqref{eq:bos_epsilon} and \eqref{eq:bos_disp}. 
The numerical results were computed from the density field on the $x$-$y$ plane passing through the origin $O$, where the integration interval in the $z$-direction was set to $-\Delta l_b < z < \Delta l_b$.  
The dashed black lines denote the inner radius $R_{\rm in}$ and the droplet radius $R_{\rm drop}$. 
The shock wave propagates from left to right for $t$  = 5.0--7.9~µs; after $t$  = 9.9~µs, the reflected shock wave travels in the opposite direction.

First, the propagation characteristics of the shock waves are examined based on the BOS measurements.
At $t$ = 4.6~µs, a negative density gradient is detected above the upper-left region outside the droplet.
At $t$~=~5.3~µs, both positive and negative density gradients are observed near the left-hand side of the droplet, while a positive density gradient appears near the droplet center.
The distance between the density gradient at $t$ = 4.6~µs and that near the droplet center at $t$ = 5.3~µs is approximately 0.8 mm, corresponding to a propagation speed of 1627.5 $\pm$ 33 m/s.
The uncertainty is primarily determined by the temporal resolution of the light source.
This propagation speed is close to the sound speed in water \cite{vogel1996shock}, indicating that the observed gradients correspond to a shock wave propagating in the surrounding fluid.
Inside the droplet, the shock wave propagates over a distance of approximately 3.3 mm between $t$ = 5.9 and 11.9~µs.
The resulting propagation speed in PFH is 503 m/s $\pm$ 33 m/s, in good agreement with the sound speed of PFH (Table~\ref{tb:parameter}).

From $t$ = 5.3 to 7.7~µs, the shock wave propagates from left to right inside the droplet. 
During this period, the shock wave converges near the point $(x,y) = (0.5 \ {\rm mm},0.0 \ {\rm mm})  $ within the droplet due to the curved liquid–liquid interface and the mismatch in the speed of sound between these two liquids.%Snell’s law.
Numerical simulations predict a focal point at $(0.5 R_{\rm drop}, 0)$, in good agreement with the BOS measurements.
% coincides with the shock wave focal point \textcolor{red}{[cite Smuele paper]}, and \textcolor{red}{the two focal points are in close agreement.}
A white region, where the density gradient is close to zero, appears between the positive and negative density gradients in the BOS measurements. 
% \textcolor{red}{As a result of focusing, the shock-induced density and its gradient both increase.}
As a result of focusing, the density at the location of the shock increases.
In this experiment, the measurable range of the integrated density gradient is limited to 24.4~kg/m$^3$~$<|\partial \rho /\partial \boldsymbol{r} |<$~494.1~kg/m$^3$ by the displacement detection technique (FCD). 
Note that the maximum limit was calculated from the maximum measurable displacement of 21.2 pixels and the displacement correction equations (Eqs.~\eqref{eq:snell}–\eqref{eq:P2P3P4}).
When the integrated density gradient exceeds the measurable upper limit, phase wrapping occurs in the displacement field, which causes the converging shock wave to display a white region between the positive and negative distributions.

At $t$  = 7.9~µs, the BOS captured the shock wave outside the droplet on the right side as a negative density gradient.
The characteristic impedance of PFH and water are 8.6~$\times$~10$^{-5}$ and 24.1~$\times$~10$^{-5}$~Pa$\cdot$s/m, respectively.
With a reflection coefficient of 0.48, both reflected and transmitted waves are expected.
The curved density gradient observed outside the droplet corresponds to a wave transmitted through the PFH–water interface.

After $t$ = 9.9~µs, the phase of the shock wave acquires an additional value of $\pi$, resulting in an inversion of the pressure amplitude compared with that before $t$ = 9.9~µs. 
This inversion cannot be attributed to reflection at the PFH–water interface, since the reflection coefficient is positive,~0.48.
Instead, it is likely associated with a Gouy phase shift, as predicted in previous studies \cite{pezeril_direct_2011,veysset_single-bubble_2018}.
The Gouy phase shift is a well-known phenomenon in optics, whereby a 3-D converging wave acquires an axial phase shift of $\pi$ upon passing through its focus during propagation from $-\infty$ to $\infty$ \cite{feng2001physical}.
Gouy phase shift has also been reported to occur for focused ultrasound in liquids \cite{lee2020origin, fiorini2024positivepressuremattersacoustic}.
More recently, Fiorini \textit{et al.,} demonstrated, through both numerical simulations and experiments, that Gouy phase shift can occur also during shock wave focusing inside droplets \cite{fiorini2026cavitation}.

The integrated density gradient field of the shock wave observed in the numerical simulation was in good agreement with the BOS measurements.
The numerical simulations further revealed the detailed integrated density gradient fields in the defocused region $R_{\rm{in}} < r < R_{\rm{drop}}$, which is over the depth of field with camera lens, as well as at $t$ = 7.3 and 7.5~µs, where the measurement was strongly affected by the limitation of FCD.
In contrast, the BOS technique, which achieved high resolution of imaging 0.45~µm/pixel, successfully quantified the linear positive density gradient located ahead of the negative density gradient field at $t$ = 9.9--11.5~µs.
A more quantitative comparison is provided in Section~\ref{sec:result-densitygradient_line} and Fig.~\ref{fig:result-density-line}.
Therefore, these results demonstrate that BOS can provide integrated density gradient fields within its measurement limits and enable spatiotemporal quantitative measurements of shock waves propagating inside droplets without calibration.

%---------------------------------------------
\clearpage
%-------------------------------------------

%---------------------------------------
\subsubsection{Axial integrated density gradient profile}\label{sec:result-densitygradient_line}
%-------------------------------------------
\begin{figure}[ht]
    \centering
    \includegraphics[width=0.4\linewidth]{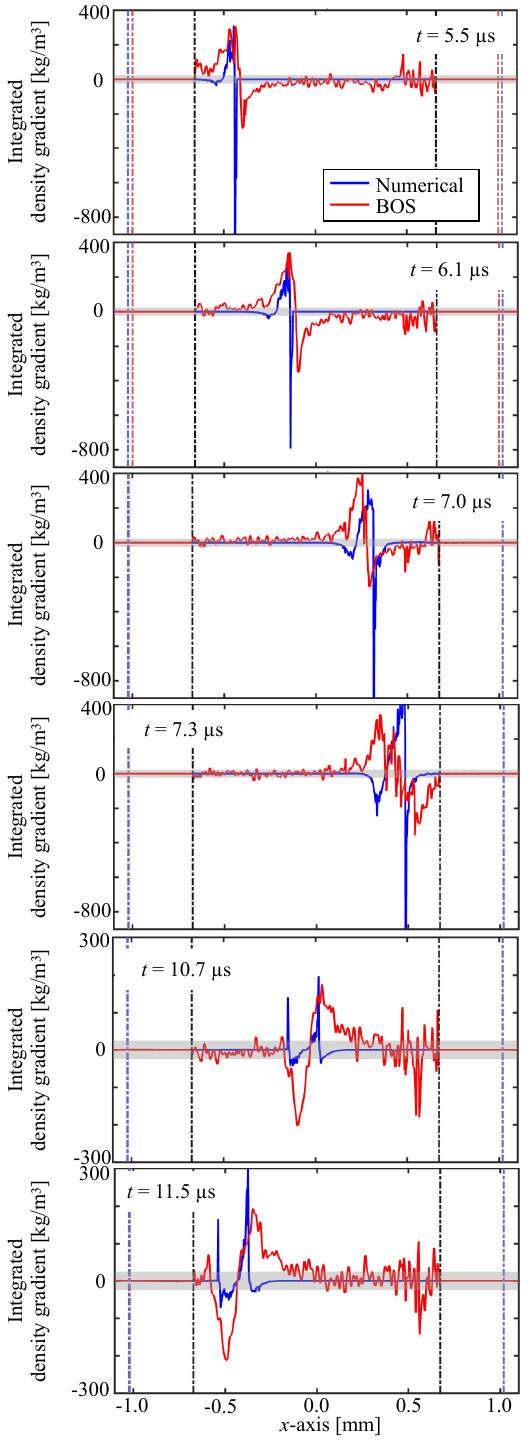}
    \caption{The $x$-direction density gradient, integrated along the $z$-axis at $t$ = 5.5, 6.1, 7.0, 7.3, 10.7, and 11.5~µs.
    $x$ = 0 indicates the center of the droplet. 
    The solid red and blue line show the BOS measurement result and numerical result, respectively.
    The dashed red and blue lines indicate the radius $R_{\rm drop}$ in the BOS measurements and numerical results, respectively.
    The black dashed line represents the internal radius of the droplet $R_{\rm in}$, where ray tracing is effectively applied.
    The gray region indicates the lower detectable value of the BOS in this experiment, $\pm$~24.4~kg/m$^3$, which was estimated from the uncertainty of the deviation angle.}
    \label{fig:result-density-line}
\end{figure}
%--------------------------------------------

The experimental and numerical density gradients in the $x$-direction, integrated along the $z$-axis, are compared in Fig.~\ref{fig:result-density-line}.
% The experimental and numerical $x$-direction density gradients, integrated along the $z$-axis are compared in Fig.~\ref{fig:result-density-line}.
The results at $t$ = 5.5~µs and $t$ = 6.1~µs show the shock-wave behavior immediately after transmission from the water into the droplet. 
At both times, the integrated values obtained from BOS and the numerical simulation exhibit strong negative gradients, and the overall trends of the two data sets are in good agreement. 
However, the measured minimum value is larger than those predicted numerically. 
The numerical results exhibit a negative tail behind the peak, whereas the BOS measurement shows a smoother gradient  that does not cross the $x$-axis, as illustrated in Fig.~\ref{fig:result-density-line} for $t$~=~5.5--7.3~µs.
From the uncertainty of the deviation angle $\varepsilon_y \pm 2.9\times10^{-3}$, and the BOS definition equations (Eqs.~\eqref{eq:bos_epsilon} and \eqref{eq:bos_disp}), the lowest detectable integrated density gradient value in this experiment can be estimated and corresponds to $\pm$ 24.4~kg/m$^3$.
The gray region in Fig. \ref{fig:result-density-line} represents this detection lower limit.
At $t=$ 5.5 and 6.1~µs, the negative tail predicted by the numerical simulation falls within this limit.
% \textcolor{red}{From Section~\ref{sec:result-densitygradient_field}, the detectable minimum value of BOS in this experiment is 24.4~kg/m$^3$, while the negative tail observed at both times is approximately 20~kg/m$^3$. }
In the numerical simulation, the initial conditions for the shock wave were estimated using a Friedlander function fitted to the experimental pressure profile obtained from BOS measurements (Eq.~ \eqref{eq:friedlanderEquation}), where a negative tail is present behind the shock front. 
Moroeover, the numerical simulation solver uses a relaxation hyperbolic model that neglects energy dissipation and attenuation.
These assumptions, together with experimental limitations, are therefore considered the primary causes of the observed discrepancies.

The results at $t$ = 7.0~µs and $t$ = 7.3~µs describe the shock wave focusing within the droplet. 
A distance of about 0.1 mm is observed between the coordinates of the experimental and numerical maxima.
The red and blue dashed lines indicate the droplet radius do not coincide at, for example, $t$ = 5.5 and 6.1~µs.
Experimentally, the time-series data were acquired via repeated single-frame acquisitions using a synchronization technique, with three droplets of radii around 0.91, 1.00, and 1.02~mm. 
In contrast, simulations that neglect dissipation predict higher pressures at the focus than those observed experimentally, which locally accelerate the propagation speed near the focus. 
These findings suggest that the observed discrepancy likely results from variations in droplet radius and the conditions employed in the numerical simulations.

%----------------------------------------------------
\begin{figure}[hb]
    \centering
    \includegraphics[width=1.0\linewidth]{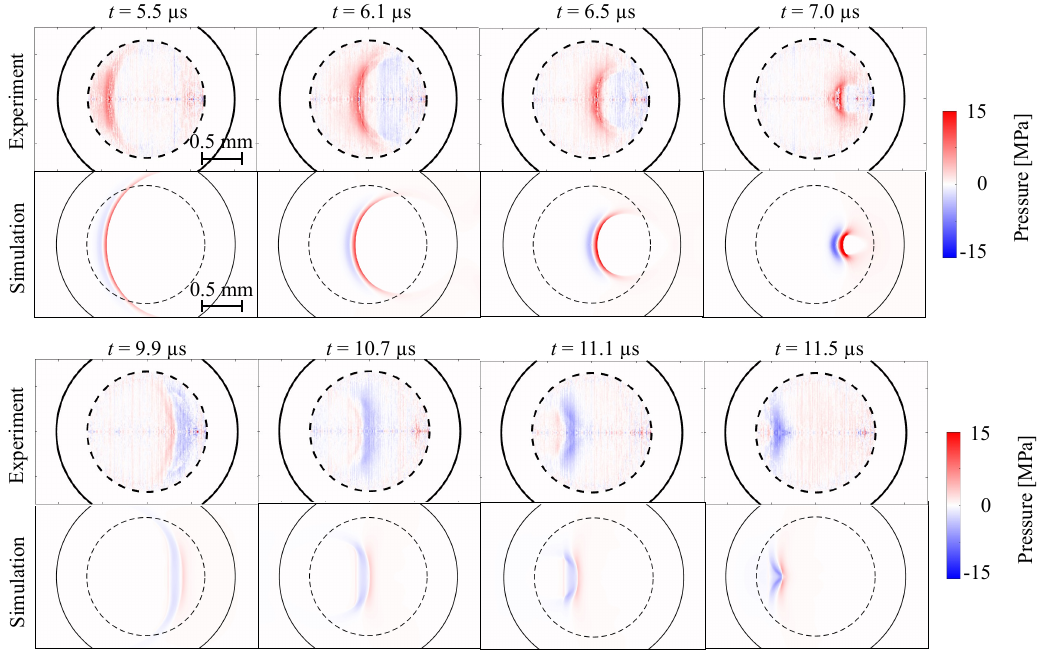}
    \caption{Pressure field on the $x$–$y$ plane passing through the origin $o$.
    The top panels show the fields measured by BOS, while the bottom panels present those obtained from the ECOGEN simulations.
    }
    \label{fig:result-p-field}
\end{figure}
%---------------------------------------------------

The results at $t$ = 10.7~µs and 11.5~µs show the behavior of the shock wave reflected at the PFH–water interface. 
Both the experimental measurements and the numerical simulations exhibit a reduced absolute amplitude compared with the result at $t =$~6.1~µs, which is attributed to shock wave reflection at the PFH–water interface. 
However, the attenuation predicted by the numerical simulation is stronger than that observed experimentally.
Similar discrepancies between experimental and numerical results for shock wave propagation in a water column were reported by Sembian \textit{et al.} \cite{sembian2016plane}. 
Damianos \textit{et al.} suggested that such discrepancies arise from the finite time required for thermodynamic equilibrium to be re-established after shock passage in gas–liquid mixtures, where distinct thermodynamic states are induced in each phase \cite{damianos2025effect}. 
In the present numerical simulations, non-equilibrium states are assumed to relax instantaneously to equilibrium, and finite-rate thermal relaxation effects are not accounted for \cite{Kapila2001ECOGENModel}. 
%In the experiments, time-series fields were obtained from repeated measurements using a synchronized acquisition system. 
% Since the shock wave propagated through the same droplet more than several tens of times, the validity of the thermodynamic equilibrium assumption is questionable even on the experimental side. 
Since the same droplet is used for multiple data acquisition within a time span of less that a minute, the validity of the thermodynamic equilibrium assumption is questionable even on the experimental side. 
Therefore, the observed discrepancy is likely attributed to limitation in the shock wave propagation model rather than to measurement errors.
Taken together, these results indicate that BOS provides a convenient framework for assessing discrepancies between assumptions used in numerical shock-wave simulations and experimentally observed behavior.

% The proposed BOS technique enables robust quantitative spatiotemporal measurements of shock waves inside a spherical droplet without calibration.
% \textcolor{blue}{Schlieren imaging, shadowgraphy, and holographic interferometry require calibration experiments to obtain quantitative fields.
% By contrast, the proposed BOS technique enables robust quantitative spatiotemporal measurements of shock waves inside a spherical droplet without calibration.
% These results indicate that BOS provides a convenient framework for assessing discrepancies between assumptions used in numerical shock-wave simulations and experimentally observed behavior.
% The BOS setup is experimentally simple and is applicable not only to shock-wave dynamics in droplets but generally also to measurements of spatially complex waveforms and nonlinear wave propagation in the presence of curved interfaces.}

\subsection{Spatiotemporal pressure}\label{sec:result-pressure}
\subsubsection{Pressure field}

We compare the spatiotemporal pressure fields obtained from the BOS measurements and numerical simulations using Fig.~\ref{fig:result-p-field}.
The top panels show the pressure fields measured by the BOS technique, whereas the bottom panels present those computed by the numerical simulations.
The solid black line denotes the droplet radius $R_{\rm drop}$, while the dashed black line indicates the inner radius $R_{\rm in}$.
At all times, the BOS and simulation results exhibit broadly similar pressure amplitudes and spatial distributions.
In particular, the BOS successfully captures the phase shift of the shock wave reflected inside the droplet, which happens between at $t$ = 7.0~µs and $t$ = 9.9~µs.

In the BOS results, the absence of a negative tail behind the shock wave and the appearance of a negative pressure region ahead of the shock front at $t$ = 6.1~µs and 6.5~µs differ significantly from the numerical results.
The discrepancy in the negative tail arises from the criteria of the BOS and from the assumptions used in the numerical simulations, as discussed in section \ref{sec:result-densitygradient}.
The negative pressure observed ahead of the shock front is attributed to the defocus occurring in the region $R_{\rm in} \le r \le R_{\rm drop}$
When reconstructing density from the density gradient in BOS, the integration used $R_{\rm in}$ as the boundary, where the density is considered constant and equal to PFH density $\rho_0$, as shown in Fig.~\ref{fig:exp-analysis_condition}.
In contrast, the pressure field obtained from numerical simulations indicates that the pressure is larger than atmospheric pressure at the boundary.
As a result, the density at $R_{\rm in}$ is actually higher than the assumed PFH density $\rho_0$.
Therefore, the pressure estimated from BOS leads to an underestimation than that computed by numerical simulations through the integration process.

\subsubsection{Axial pressure profile}
%----------------------------------------------------
\begin{figure}[ht]
    \centering
    \includegraphics[width=0.5\linewidth]{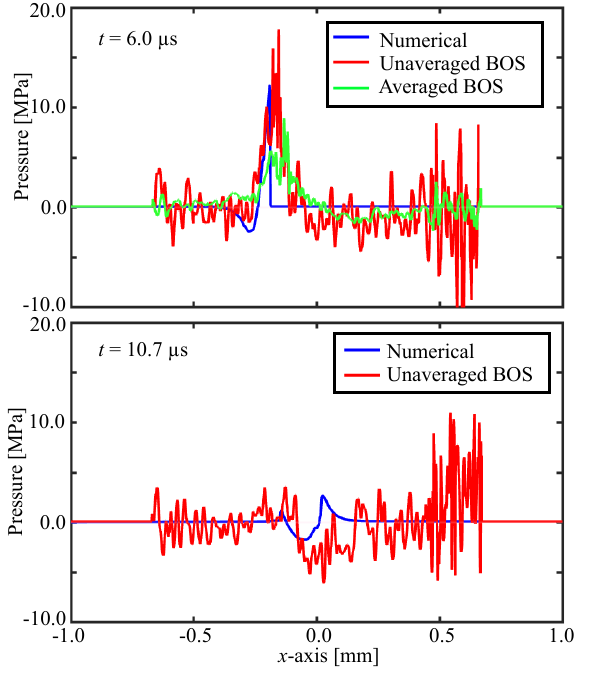}
    \caption{Pressure profiles along the $y$-axis at $t = 6.0$~µs and $t =$ 10.7~µs. The red and blue lines represent the BOS measurements and the numerical simulations, respectively. The green line denotes the BOS result obtained by averaging ten deviation angle fields.}
    \label{fig:result-pressure-line}
\end{figure}
%----------------------------------------------------
The experimentally and numerically obtained pressure along the $y$-axis are compared in Fig.~\ref{fig:result-pressure-line}.
The green line shows the BOS measurement result using averaged ten deviation angle fields.
The top graph shows the result at 6.0~µs (during focusing), while the bottom graph shows the result at 10.7~µs (after reflection).

At $t$ = 6.0~µs, when the shock wave propagated in the positive $x$-direction, a positive pressure wave was obtained. 
In contrast, at $t$ = 10.7~µs, which corresponds to the shock wave reflected at the PFH–water interface, a negative pressure wave was observed. 
% Notably, at $t$ = 6.0~µs, the peak pressure obtained from the numerical simulation, 12.9 MPa, was in close agreement with that measured using the BOS technique, 10.0 $\pm 4.9$~MPa. 
Notably, at $t$ = 6.0~µs, the peak pressure obtained by the numerical simulation (12.2~MPa) agrees well with that estimated from a single BOS dataset (17.8~$\pm~3.5$~MPa).
In the numerical results, a negative tail appeared behind the shock front at $t$ = 6.0~µs, whereas a positive tail was observed at $t$ = 10.7~µs.
These tails were not quantified in the BOS measurements. 
As discussed in Section \ref{sec:result-densitygradient}, this discrepancy likely arises primarily from two factors: the measurement limitations of the BOS and the simplifying assumptions adopted in the numerical simulations.

The uncertainty of the pressure reconstructed from a single BOS dataset reaches $\pm$~3.5~MPa (red line), with larger uncertainties near the symmetry axis ($y = 0$) due to the reconstruction procedure (Fig.~\ref{fig:exp-uncertainty}). 
At $t$ = 6.0~µs, averaging ten deviation angle fields reduces the uncertainty to $\pm$~0.3~MPa (green line), demonstrating that averaging effectively suppresses random noise.
However, the resulting peak pressure from the averaged BOS data (9.0~$\pm$~0.3~MPa) remains lower than both the unaveraged value (17.8~$\pm~3.5$~MPa) and the numerical simulation result (12.2~MPa).
The hydrophone measurement in the absence of the droplet exhibits an uncertainty of $\pm$~4.8~MPa, indicating that the laser pulse energy cannot be assumed strictly constant during repeated measurements.
When the uncertainty in the laser energy is taken into account, the discrepancy between the numerical simulation results and the averaged and unaveraged BOS results is nearly negligible.
% In addition, the present BOS configuration suffers from defocus effects in the region $R_{\rm in} < r < R_{\rm drop}$, and the integration procedure using $R_{\rm in}$ as the inner boundary likely leads to an underestimation of density and pressure.
% The reduced peak pressure obtained after averaging is therefore attributed primarily to these experimental limitations.

% These results demonstrate that BOS enables non-intrusive measurements of spatiotemporally evolving, high-amplitude pressure waves without disturbing the flow field inside a spherical object. 
% Repeated measurements under identical conditions can further reduce the measurement uncertainty of the proposed BOS technique.
% According to the BOS formulation (Eq.~\ref{eq:bos_disp_epsilon}), the displacement depends on the camera's magnification and the distance $l_b$. 
% The measurable upper pressure limit can therefore be increased by employing a high spatial resolution camera (e.g. 1.0~$\times 10^{8}$~pixels) or by reducing $l_b$. 
% Conversely, increasing $l_b$ enhances the sensitivity to small displacements, thereby lowering the minimum detectable pressure.
The present results demonstrate that, by applying the reconstruction procedure and EoS to the integrated density gradient field, BOS enables non-intrusive measurements of high-amplitude pressure waves without disturbing the flow field inside a spherical object.
Repeated measurement under identical conditions can further reduce the measurement uncertainty of the proposed BOS technique.
Furthermore, the BOS method has the advantage that the measurement limits (24.4~kg/m$^3$~$<|\partial \rho /\partial \boldsymbol{r} |<$~494.1~kg/m$^3$) can be adjusted by modifying the experimental conditions.
According to the BOS formulation (Eq.~\ref{eq:bos_disp_epsilon}), the displacement depends on the camera magnification and the distance $l_b$.
The measurable upper pressure limit can therefore be increased by employing a high spatial resolution camera (e.g. 1.0~$\times 10^{8}$~pixels) or by reducing $l_b$.
Conversely, increasing $l_b$ enhances the sensitivity to small displacements, thereby lowering the minimum detectable pressure.

% %%===============================================
\section{Conclusion}\label{sec:conclusion}

We have proposed a background-oriented schlieren (BOS) technique for the quantitative spatiotemporal measurement of shock wave–droplet interactions, employing a novel ray-tracing correction, a synchronization technique, and a projected background. 
The measurements capture laser-induced shock waves propagating inside and outside a 1-mm-radius perfluorohexane (PFH) droplet placed on an agarose substrate and immersed in water.
The experimentally quantified time-evolving integrated density gradient and reconstructed pressure fields have been successfully compared with numerical simulations to evaluate the accuracy of the measurement. 

The BOS measurements—including the density and pressure distributions, the sound speeds outside and inside the droplet (1627.5 $\pm$ 33 m/s and 503 $\pm$ 33 m/s, respectively), the location of the shock‐wave focusing ($(x,y)$ = (0.5 mm, 0.0 mm)), and the maximum pressure (17.8 $\pm$ 3.5 MPa)—are in close agreement with the numerical results. 
Notably, the BOS technique successfully measures the Gouy phase shift occurring during shock focusing, as well as the the attenuation effect resulting from repeated experiments, both of which had previously been hypothesized \cite{fiorini2025positive,damianos2025effect}.

Conventional techniques, such as schlieren imaging and shadowgraphy, require calibration experiments to estimate quantitative fields from image intensity. 
The proposed BOS technique enables directly quantitative measurements of high-speed multiphase flow phenomena both inside and outside spherical interfaces without calibration. The measurement limits (24.4~kg/m$^3$~$<|\partial \rho /\partial \boldsymbol{r} |<$~494.1~kg/m$^3$) BOS are determined by the spatial resolution of the image sensor, the checker size of the background pattern, and the distance $l_b$, and are therefore adjustable.
Consequently, the technique can be further applied to capture spatially complex acoustic waveforms and nonlinear wave propagation~\cite{shpak2014acoustic}, to evaporating droplets~\cite{bell2022concentration}, as well as to acoustically driven droplets \cite{liu2025acousto,koroyasu2023microfluidic}.
With further improvements, the proposed BOS technique may advance the understanding of multiphase flow phenomena and find applications in agricultural sprays, aircraft propulsion systems, drug delivery systems, and rainfall processes.
%----------------------------------------------------
\section*{Acknowledgement}
The authors acknowledge the financial support from the Swiss National Science Foundation (Grant No.\ 200567), ETH Zürich, the Tokyo University of Agriculture and Technology, and the Japan Society for the Promotion of Science KAKENHI (Grant Nos.\ JP22KJ1239, JP24H00289); the Japan Science and Technology Agency PRESTO (Grant No.\ JPMJPR21O5); and the Japan Agency for Medical Research and Development (Grant No.\ JP22he0422016).
The authors also thank Dr.~Guillaume T. Bokman and Mr.~Anunay Prasanna for their assistance in the development of the numerical simulations and for insightful discussions.

\bibliographystyle{ieeetr}
\bibliography{ref}

@article{Prasanna2025PFHEquationOfState,
    author = {Prasanna, A. and Bokman, G. T. and Fiorini, S. and Sieber, A. and Lukić, B. and Foster, D. and Supponen, O.},
    title = {Shock-compression-based equation of state for perfluorohexane},
    journal = {Physical Review E},
    volume = {112},
    issue = {6},
    year = {2025},
    month = {December} 
}

@article{Kapila2001ECOGENModel,
    author = {Kapila, A.K. and Menikoff, R. and Bdzil J. D. and Son, S. F. and Stewart, D.S.},
    title = {Two-phase modeling of deflagration-to-detonation transition in granular materials: Reduced equations},
    journal = {Physics of Fluids},
    volume = {13},
    issue = {10},
    pages = {3002-3024},
    year = {2001},
    month = {January} 
}

@article{Saurel2009ECOGENModel,
    author = {Saurel, R. and Petitpas, F.},
    title = {Simple and efficient relaxation methods for interfaces separating compressible fluids, cavitating flows and shocks in multiphase mixtures},
    journal = {Journal of Computational Physics},
    volume = {228},
    issue = {5},
    pages = {1678-1712},
    year = {2009},
    month = {March} 
}

@article{ichihara2022background,
  title={Background-oriented schlieren technique with vector tomography for measurement of axisymmetric pressure fields of laser-induced underwater shock waves},
  author={Ichihara, Sayaka and Shimazaki, Takaaki and Tagawa, Yoshiyuki},
  journal={Experiments in Fluids},
  volume={63},
  number={11},
  pages={182},
  year={2022},
  publisher={Springer}
}

@misc{fiorini2024positivepressuremattersacoustic,
      title={Positive pressure matters in acoustic droplet vaporization}, 
      author={Samuele Fiorini and Anunay Prasanna and Gazendra Shakya and Marco Cattaneo and Outi Supponen},
      year={2024},
      eprint={2407.16455},
      archivePrefix={arXiv},
      primaryClass={physics.flu-dyn},
      url={https://arxiv.org/abs/2407.16455}, 
}

@article{venkatakrishnan2005density,
  title={Density measurements in an axisymmetric underexpanded jet by background-oriented schlieren technique},
  author={Venkatakrishnan, Lakshmi},
  journal={AIAA journal},
  volume={43},
  number={7},
  pages={1574--1579},
  year={2005}
}

@article{lee2020origin,
  title={Origin of Gouy phase shift identified by laser-generated focused ultrasound},
  author={Lee, Taehwa and Cheong, Yeonjoon and Baac, Hyoung Won and Guo, L Jay},
  journal={ACS Photonics},
  volume={7},
  number={11},
  pages={3236--3245},
  year={2020},
  publisher={ACS Publications}
}

@article{gladstone1863xiv,
  title={XIV. Researches on the refraction, dispersion, and sensitiveness of liquids},
  author={Gladstone, John Hall and Dale, Thomas P},
  journal={Philosophical Transactions of the Royal Society of London},
  number={153},
  pages={317--343},
  year={1863},
  publisher={The Royal Society London}
}

@article{shpak2014acoustic,
  title={Acoustic droplet vaporization is initiated by superharmonic focusing},
  author={Shpak, Oleksandr and Verweij, Martin and Vos, Hendrik J and de Jong, Nico and Lohse, Detlef and Versluis, Michel},
  journal={Proceedings of the National Academy of Sciences},
  volume={111},
  number={5},
  pages={1697--1702},
  year={2014},
  publisher={National Acad Sciences}
}

@article{bokman_scaling_2023,
	title = {Scaling laws for bubble collapse driven by an impulsive shock wave},
	volume = {967},
	issn = {0022-1120, 1469-7645},
	url = {https://www.cambridge.org/core/product/identifier/S0022112023005141/type/journal_article},
	doi = {10.1017/jfm.2023.514},
	language = {en},
	urldate = {2024-11-14},
	journal = {Journal of Fluid Mechanics},
	author = {Bokman, Guillaume T. and Biasiori-Poulanges, Luc and Meyer, Daniel W. and Supponen, Outi},
	month = jul,
	year = {2023},
	pages = {A33},
}

@article{le_metayer_elaboration_2004,
	title = {Élaboration des lois d'état d'un liquide et de sa vapeur pour les modèles d'écoulements diphasiques},
	volume = {43},
	copyright = {https://www.elsevier.com/tdm/userlicense/1.0/},
	issn = {12900729},
	url = {https://linkinghub.elsevier.com/retrieve/pii/S1290072903001443},
	doi = {10.1016/j.ijthermalsci.2003.09.002},
	language = {fr},
	number = {3},
	urldate = {2024-11-14},
	journal = {International Journal of Thermal Sciences},
	author = {Le Métayer, O and Massoni, J and Saurel, R},
	month = mar,
	year = {2004},
	pages = {265--276},
}

@article{schmidmayer_ecogen_2020,
	title = {{ECOGEN}: {An} open-source tool for multiphase, compressible, multiphysics flows},
	volume = {251},
	issn = {00104655},
	shorttitle = {{ECOGEN}},
	url = {https://linkinghub.elsevier.com/retrieve/pii/S0010465519303959},
	doi = {10.1016/j.cpc.2019.107093},
	language = {en},
	urldate = {2024-11-14},
	journal = {Computer Physics Communications},
	author = {Schmidmayer, Kevin and Petitpas, Fabien and Le Martelot, S\'ebastien and Daniel, \'Eric},
	month = jun,
	year = {2020},
	pages = {107093},
}

@article{pishchalnikov_high-speed_2019,
	title = {High-speed video microscopy and numerical modeling of bubble dynamics near a surface of urinary stone},
	volume = {146},
	issn = {0001-4966, 1520-8524},
	url = {https://pubs.aip.org/jasa/article/146/1/516/994511/High-speed-video-microscopy-and-numerical-modeling},
	doi = {10.1121/1.5116693},
	language = {en},
	number = {1},
	urldate = {2024-11-14},
	journal = {The Journal of the Acoustical Society of America},
	author = {Pishchalnikov, Yuri A. and Behnke-Parks, William M. and Schmidmayer, Kevin and Maeda, Kazuki and Colonius, Tim and Kenny, Thomas W. and Laser, Daniel J.},
	month = jul,
	year = {2019},
	pages = {516--531},
}

@article{dorschner_formation_2020,
	title = {On the formation and recurrent shedding of ligaments in droplet aerobreakup},
	volume = {904},
	issn = {0022-1120, 1469-7645},
	url = {https://www.cambridge.org/core/product/identifier/S0022112020006990/type/journal_article},
	doi = {10.1017/jfm.2020.699},
	language = {en},
	urldate = {2024-11-14},
	journal = {Journal of Fluid Mechanics},
	author = {Dorschner, Benedikt and Biasiori-Poulanges, Luc and Schmidmayer, Kevin and El-Rabii, Hazem and Colonius, Tim},
	month = dec,
	year = {2020},
	pages = {A20},
}

@article{venkatakrishnan2004density,
  title={Density measurements using the background oriented schlieren technique},
  author={Venkatakrishnan, L and Meier, GEA},
  journal={Experiments in Fluids},
  volume={37},
  pages={237--247},
  year={2004},
  publisher={Springer}
}

@article{raffel2015background,
  title={Background-oriented schlieren (BOS) techniques},
  author={Raffel, Markus},
  journal={Experiments in Fluids},
  volume={56},
  pages={1--17},
  year={2015},
  publisher={Springer}
}

@article{ota2015improvement,
  title={Improvement in spatial resolution of background-oriented schlieren technique by introducing a telecentric optical system and its application to supersonic flow},
  author={Ota, Masanori and Leopold, Friedrich and Noda, Ryusuke and Maeno, Kazuo},
  journal={Experiments in fluids},
  volume={56},
  pages={1--10},
  year={2015},
  publisher={Springer}
}

@inproceedings{leopold2012increase,
  title={Increase of accuracy for CBOS by background projection},
  author={Leopold, F and Jagusinski, F and Demeautis, C and Ota, M and Klatt, D},
  booktitle={Proceedings of 15th international symposium on flow visualization (ISFV15-087), Minsk, Belarus},
  pages={23--52},
  year={2012}
}

@article{ichihara2025high,
  title={High-resolution pressure imaging via background-oriented schlieren tomography: A spatiotemporal measurement for MHz ultrasound fields and hydrophone calibration},
  author={Ichihara, Sayaka and Yamagishi, Masato and Kurashina, Yuta and Ota, Masanori and Tagawa, Yoshiyuki},
  journal={Ultrasonics},
  pages={107614},
  year={2025},
  publisher={Elsevier}
}

@article{meier2002computerized,
  title={Computerized background-oriented schlieren},
  author={Meier, GEA},
  journal={Experiments in fluids},
  volume={33},
  number={1},
  pages={181--187},
  year={2002},
  publisher={Springer}
}

@article{shimazaki2022background,
  title={Background oriented schlieren technique with fast Fourier demodulation for measuring large density-gradient fields of fluids},
  author={Shimazaki, Takaaki and Ichihara, Sayaka and Tagawa, Yoshiyuki},
  journal={Experimental Thermal and Fluid Science},
  volume={134},
  pages={110598},
  year={2022},
  publisher={Elsevier}
}

@article{kang2004quantitative,
  title={Quantitative visualization of flow inside an evaporating droplet using the ray tracing method},
  author={Kang, Kwan Hyoung and Lee, Sang Joon and Lee, Choung Mook and Kang, In Seok},
  journal={Measurement Science and Technology},
  volume={15},
  number={6},
  pages={1104},
  year={2004},
  publisher={IOP Publishing}
}

@article{hecht2002optics,
  title={Optics, ed},
  author={Hecht, Eugene},
  journal={New York: Addison \& Wesley},
  year={2002}
}

@article{tripathi2022interactions,
  title={Interactions between shock waves and liquid droplet clusters: Interfacial physics},
  author={Tripathi, Mitansh and Ganti, Himakar and Khare, Prashant},
  journal={Journal of Fluids Engineering},
  volume={144},
  number={10},
  pages={101401},
  year={2022},
  publisher={American Society of Mechanical Engineers}
}

@article{forehand2023numerical,
  title={A numerical assessment of shock--droplet interaction modeling including cavitation},
  author={Forehand, RW and Nguyen, KC and Anderson, CJ and Shannon, R and Grace, SM and Kinzel, MP},
  journal={Physics of Fluids},
  volume={35},
  number={2},
  year={2023},
  publisher={AIP Publishing}
}

@article{feng2001physical,
  title={Physical origin of the Gouy phase shift},
  author={Feng, Simin and Winful, Herbert G},
  journal={Optics letters},
  volume={26},
  number={8},
  pages={485--487},
  year={2001},
  publisher={Optica Publishing Group}
}

@book{brujan2010cavitation,
  title={Cavitation in Non-Newtonian fluids: with biomedical and bioengineering applications},
  author={Brujan, Emil},
  year={2010},
  publisher={Springer Science \& Business Media}
}

@article{hayasaka2016optical,
  title={Optical-flow-based background-oriented schlieren technique for measuring a laser-induced underwater shock wave},
  author={Hayasaka, Keisuke and Tagawa, Yoshiyuki and Liu, Tianshu and Kameda, Masaharu},
  journal={Experiments in Fluids},
  volume={57},
  pages={1--11},
  year={2016},
  publisher={Springer}
}

@article{westerweel1997fundamentals,
  title={Fundamentals of digital particle image velocimetry},
  author={Westerweel, Jerry},
  journal={Measurement science and technology},
  volume={8},
  number={12},
  pages={1379},
  year={1997},
  publisher={IOP Publishing}
}

@article{atcheson2009evaluation,
  title={An evaluation of optical flow algorithms for background oriented schlieren imaging},
  author={Atcheson, Bradley and Heidrich, Wolfgang and Ihrke, Ivo},
  journal={Experiments in fluids},
  volume={46},
  pages={467--476},
  year={2009},
  publisher={Springer}
}

@article{wildeman2018real,
  title={Real-time quantitative Schlieren imaging by fast Fourier demodulation of a checkered backdrop},
  author={Wildeman, Sander},
  journal={Experiments in Fluids},
  volume={59},
  number={6},
  pages={97},
  year={2018},
  publisher={Springer}
}

@book{glassner1989introduction,
  title={An introduction to ray tracing},
  author={Glassner, Andrew S},
  year={1989},
  publisher={Morgan Kaufmann}
}

@article{supponen2017shock,
  title={Shock waves from nonspherical cavitation bubbles},
  author={Supponen, Outi and Obreschkow, Danail and Kobel, Philippe and Tinguely, Marc and Dorsaz, Nicolas and Farhat, Mohamed},
  journal={Physical Review Fluids},
  volume={2},
  number={9},
  pages={093601},
  year={2017},
  publisher={APS}
}

@article{vogel1996shock,
  title={Shock wave emission and cavitation bubble generation by picosecond and nanosecond optical breakdown in water},
  author={Vogel, Alfred and Busch, S and Parlitz, U},
  journal={The Journal of the Acoustical Society of America},
  volume={100},
  number={1},
  pages={148--165},
  year={1996},
  publisher={Acoustical Society of America}
}

@article{van2014density,
  title={Density measurements using near-field background-oriented Schlieren},
  author={van Hinsberg, Nils P and R{\"o}sgen, Thomas},
  journal={Experiments in fluids},
  volume={55},
  number={4},
  pages={1720},
  year={2014},
  publisher={Springer}
}

@article{yamamoto2022contactless,
  title={Contactless pressure measurement of an underwater shock wave in a microtube using a high-resolution background-oriented schlieren technique},
  author={Yamamoto, Shota and Shimazaki, Takaaki and Franco-G{\'o}mez, Andr{\'e}s and Ichihara, Sayaka and Yee, Jingzu and Tagawa, Yoshiyuki},
  journal={Experiments in Fluids},
  volume={63},
  number={9},
  pages={142},
  year={2022},
  publisher={Springer}
}

@article{hsiang1995drop,
  title={Drop deformation and breakup due to shock wave and steady disturbances},
  author={Hsiang, L-P and Faeth, Gerard M},
  journal={International Journal of Multiphase Flow},
  volume={21},
  number={4},
  pages={545--560},
  year={1995},
  publisher={Elsevier}
}

@book{settles2001schlieren,
  title={Schlieren and shadowgraph techniques: visualizing phenomena in transparent media},
  author={Settles, Gary S},
  year={2001},
  publisher={Springer Science \& Business Media}
}

@article{shinjo2011surface,
  title={Surface instability and primary atomization characteristics of straight liquid jet sprays},
  author={Shinjo, J and Umemura, A},
  journal={International Journal of Multiphase Flow},
  volume={37},
  number={10},
  pages={1294--1304},
  year={2011},
  publisher={Elsevier}
}

@article{sharma2021shock,
  title={Shock induced aerobreakup of a droplet},
  author={Sharma, Shubham and Singh, Awanish Pratap and Rao, S Srinivas and Kumar, Aloke and Basu, Saptarshi},
  journal={Journal of Fluid Mechanics},
  volume={929},
  pages={A27},
  year={2021},
  publisher={Cambridge University Press}
}

@article{sembian2016plane,
  title={Plane shock wave interaction with a cylindrical water column},
  author={Sembian, Sundarapandian and Liverts, Michael and Tillmark, Nils and Apazidis, Nicholas},
  journal={Physics of Fluids},
  volume={28},
  number={5},
  year={2016},
  publisher={AIP Publishing}
}

@article{haas1987interaction,
  title={Interaction of weak shock waves with cylindrical and spherical gas inhomogeneities},
  author={Haas, J-F and Sturtevant, Bradford},
  journal={Journal of Fluid Mechanics},
  volume={181},
  pages={41--76},
  year={1987},
  publisher={Cambridge University Press}
}

@article{igra2010shock,
  title={Shock-water column interaction, from initial impact to fragmentation onset},
  author={Igra, D and Sun, M},
  journal={AIAA journal},
  volume={48},
  number={12},
  pages={2763--2771},
  year={2010}
}

@article{igra2001investigation,
  title={Investigation of aerodynamic breakup of a cylindrical water droplet},
  author={Igra, D and Takayama, K},
  journal={Atomization and Sprays},
  volume={11},
  number={2},
  year={2001},
  publisher={Begel House Inc.}
}

@article{igra2003experimental,
  title={Experimental investigation of two cylindrical water columns subjected to planar shock wave loading},
  author={Igra, D and Takayama, Kazuyoshi},
  journal={Journal of fluids engineering},
  volume={125},
  number={2},
  pages={325--331},
  year={2003},
  publisher={American Society of Mechanical Engineers Digital Collection}
}

@article{field1989effects,
  title={The effects of target compliance on liquid drop impact},
  author={Field, JE and Dear, JP and Ogren, JE},
  journal={Journal of Applied Physics},
  volume={65},
  number={2},
  pages={533--540},
  year={1989},
  publisher={American Institute of Physics}
}

@article{damianos2025effect,
  title={Effect of gas nuclei on the primary stage of shock--droplet interaction},
  author={Damianos, Sotirios and Papoutsakis, Andreas and Karathanassis, Ioannis K and Gavaises, Manolis},
  journal={International Journal of Multiphase Flow},
  pages={105478},
  year={2025},
  publisher={Elsevier}
}

@article{kinoshita2007three,
  title={Three-dimensional measurement and visualization of internal flow of a moving droplet using confocal micro-PIV},
  author={Kinoshita, Haruyuki and Kaneda, Shohei and Fujii, Teruo and Oshima, Marie},
  journal={Lab on a Chip},
  volume={7},
  number={3},
  pages={338--346},
  year={2007},
  publisher={Royal Society of Chemistry}
}

@article{liu2017micro,
  title={Micro-PIV investigation of the internal flow transitions inside droplets traveling in a rectangular microchannel},
  author={Liu, Zhaomiao and Zhang, Longxiang and Pang, Yan and Wang, Xiang and Li, Mengqi},
  journal={Microfluidics and Nanofluidics},
  volume={21},
  number={12},
  pages={180},
  year={2017},
  publisher={Springer}
}

@article{kumar2017internal,
  title={Internal flow measurements of drop impacting a solid surface},
  author={Kumar, S Santosh and Karn, Ashish and Arndt, Roger EA and Hong, Jiarong},
  journal={Experiments in Fluids},
  volume={58},
  number={3},
  pages={12},
  year={2017},
  publisher={Springer}
}

@article{miessner2020mupiv,
  title={$\mu$PIV measurement of the 3D velocity distribution of Taylor droplets moving in a square horizontal channel},
  author={Mie{\ss}ner, Ulrich and Helmers, Thorben and Lindken, Ralph and Westerweel, Jerry},
  journal={Experiments in Fluids},
  volume={61},
  number={5},
  pages={125},
  year={2020},
  publisher={Springer}
}

@article{timgren2008application,
  title={Application of the PIV technique to measurements around and inside a forming drop in a liquid--liquid system},
  author={Timgren, Anna and Tr{\"a}g{\aa}rdh, Gun and Tr{\"a}g{\aa}rdh, Christian},
  journal={Experiments in Fluids},
  volume={44},
  number={4},
  pages={565--575},
  year={2008},
  publisher={Springer}
}

@article{yamamoto2008internal,
  title={Internal flow of acoustically levitated droplet},
  author={Yamamoto, Yuji and Abe, Yutaka and Fujiwara, Akiko and Hasegawa, Koji and Aoki, Kazuyoshi},
  journal={Microgravity Science and Technology},
  volume={20},
  number={3},
  pages={277--280},
  year={2008},
  publisher={Springer}
}

@article{minor2007optical,
  title={Optical distortion correction for liquid droplet visualization using the ray tracing method: further considerations},
  author={Minor, G and Oshkai, P and Djilali, N},
  journal={Measurement Science and Technology},
  volume={18},
  number={11},
  pages={L23},
  year={2007},
  publisher={IOP Publishing}
}

@article{bell2022concentration,
  title={Concentration gradients in evaporating binary droplets probed by spatially resolved Raman and NMR spectroscopy},
  author={Bell, Alena K and Kind, Jonas and Hartmann, Maximilian and Kresse, Benjamin and Hoefler, Mark V and Straub, Benedikt B and Auernhammer, G{\"u}nter K and Vogel, Michael and Thiele, Christina M and Stark, Robert W},
  journal={Proceedings of the National Academy of Sciences},
  volume={119},
  number={15},
  pages={e2111989119},
  year={2022},
  publisher={National Academy of Sciences}
}

@article{liu2025acousto,
  title={Acousto-dewetting enables droplet microfluidics on superhydrophilic surfaces},
  author={Liu, Song and Sun, Pengcheng and Wang, Mingyue and Jiang, Yujie and Li, Jiaqi and Jia, Yuyu and Sun, Zhenhuan and Yang, Yuting and Liu, Hai and Lu, Haojian and others},
  journal={Nature Physics},
  pages={1--9},
  year={2025},
  publisher={Nature Publishing Group UK London}
}

@article{koroyasu2023microfluidic,
  title={Microfluidic platform using focused ultrasound passing through hydrophobic meshes with jump availability},
  author={Koroyasu, Yusuke and Nguyen, Thanh-Vinh and Sasaguri, Shun and Marzo, Asier and Ezcurdia, I{\~n}igo and Nagata, Yuuya and Yamamoto, Tatsuya and Nomura, Nobuhiko and Hoshi, Takayuki and Ochiai, Yoichi and others},
  journal={PNAS nexus},
  volume={2},
  number={7},
  pages={pgad207},
  year={2023},
  publisher={Oxford University Press US}
}

@article{pezeril_direct_2011,
	title = {Direct {Visualization} of {Laser}-{Driven} {Focusing} {Shock} {Waves}},
	volume = {106},
	copyright = {http://link.aps.org/licenses/aps-default-license},
	issn = {0031-9007, 1079-7114},
	url = {https://link.aps.org/doi/10.1103/PhysRevLett.106.214503},
	doi = {10.1103/PhysRevLett.106.214503},
	number = {21},
	urldate = {2025-12-16},
	journal = {Physical Review Letters},
	author = {Pezeril, T. and Saini, G. and Veysset, D. and Kooi, S. and Fidkowski, P. and Radovitzky, R. and Nelson, Keith A.},
	month = may,
	year = {2011},
	pages = {214503},
}

@article{veysset_single-bubble_2018,
	title = {Single-bubble and multibubble cavitation in water triggered by laser-driven focusing shock waves},
	volume = {97},
	issn = {2470-0045, 2470-0053},
	url = {https://link.aps.org/doi/10.1103/PhysRevE.97.053112},
	doi = {10.1103/PhysRevE.97.053112},
	number = {5},
	urldate = {2025-12-16},
	journal = {Physical Review E},
	author = {Veysset, D. and Gutiérrez-Hernández, U. and Dresselhaus-Cooper, L. and De Colle, F. and Kooi, S. and Nelson, K. A. and Quinto-Su, P. A. and Pezeril, T.},
	month = may,
	year = {2018},
	pages = {053112},
}

@article{fiorini2025positive,
  title={Positive pressure matters in acoustic droplet vaporization},
  author={Fiorini, Samuele and Prasanna, Anunay and Shakya, Gazendra and Cattaneo, Marco and Supponen, Outi},
  journal={Physical Review Research},
  volume={7},
  number={2},
  pages={023322},
  year={2025},
  publisher={APS}
}

@article{theofanous2004aerobreakup,
  title={Aerobreakup in rarefied supersonic gas flows},
  author={Theofanous, TG and Li, GJ and Dinh, Truc-Nam},
  journal={Journal of fluids engineering},
  volume={126},
  number={4},
  pages={516--527},
  year={2004},
  publisher={American Society of Mechanical Engineers Digital Collection}
}

@article{xiong2024exploration,
  title={Exploration of shock--droplet interaction based on high-fidelity simulation and improved theoretical model},
  author={Xiong, Tianheng and Shao, Changxiao and Luo, Kun},
  journal={Journal of Fluid Mechanics},
  volume={988},
  pages={A46},
  year={2024},
  publisher={Cambridge University Press}
}

@article{le2016noble,
  title={The Noble-Abel stiffened-gas equation of state},
  author={Le M{\'e}tayer, Olivier and Saurel, Richard},
  journal={Physics of Fluids},
  volume={28},
  number={4},
  year={2016},
  publisher={AIP Publishing}
}

@article{schmidmayer2019adaptive,
  title={Adaptive mesh refinement algorithm based on dual trees for cells and faces for multiphase compressible flows},
  author={Schmidmayer, Kevin and Petitpas, Fabien and Daniel, Eric},
  journal={Journal of Computational Physics},
  volume={388},
  pages={252--278},
  year={2019},
  publisher={Elsevier}
}

@article{menezes2009shock,
  title={Shock wave driven liquid microjets for drug delivery},
  author={Menezes, Viren and Kumar, Satyam and Takayama, Kazuyoshi},
  journal={Journal of Applied Physics},
  volume={106},
  number={8},
  year={2009},
  publisher={AIP Publishing}
}

@article{fiorini2026cavitation,
      title={Cavitation by phase shift of focused shock waves inside a droplet}, 
      author={Samuele Fiorini and Guillaume T. Bokman and Anunay Prasanna and Stefanos Nikolaou and Sayaka Ichihara and Bratislav Lukić and Alexander Rack and Yoshiyuki Tagawa and Outi Supponen},
      journal = {arXiv preprint 	arXiv:2603.19990},
      year={2026},
      eprint={2603.19990},
      archivePrefix={arXiv},
      primaryClass={physics.flu-dyn},
      url={https://arxiv.org/abs/2603.19990}, 
}

\appendix 
\section{Supplementary videos}\label{sec:suppMat}
The experimental videos obtained using background-oriented schlieren (BOS) techniques are available for download. Specifically, the dataset includes:
\begin{itemize}
    \item \texttt{density\_gradient\_bos.avi}, showing the density gradient fields integrated along the $z$-direction in both the $x$- and $y$-directions, as described in Section~\ref{sec:result-densitygradient}.
   
    \item \texttt{pressure\_bos.avi}, showing the pressure field on the $xy$-plane, as described in Section~\ref{sec:result-pressure}.
\end{itemize}
In each video, the time stamp displayed above the frames indicates the elapsed time $t$ relative to the instant when laser irradiation was triggered. 
In the density gradient video, the left panel shows the $x$-direction density gradient field, while the right panel shows the $y$-direction density gradient field.
\end{document}